\documentclass[sn-mathphys,Numbered]{sn-jnl}

\usepackage{graphicx}%
\usepackage{multirow}%
\usepackage{amsmath,amssymb,amsfonts,bm}%
\usepackage{amsthm}%
\usepackage{mathrsfs}%
\usepackage[title]{appendix}%
\usepackage{xcolor}%
\usepackage{textcomp}%
\usepackage{manyfoot}%
\usepackage{booktabs}%
\usepackage{algorithm}%
\usepackage{algorithmicx}%
\usepackage{algpseudocode}%
\usepackage{listings}%
\usepackage{subcaption,soul}
\theoremstyle{thmstyletwo}%

\theoremstyle{thmstylethree}%

\begin{document}

\title[Article Title]{CLEP-GAN: An Innovative Approach to Subject-Independent ECG Reconstruction from PPG Signals}


\author[1]{\fnm{Xiaoyan} \sur{Li}}\email{xiaoy.li@mail.utoronto.ca}
\author[3]{\fnm{Shixin} \sur{Xu}}\email{shixin.xu@dukekunshan.edu.cn}
\author[4]{\fnm{Faisal} \sur{Habib}}\email{fhabib@fields.utoronto.ca}
\author[2]{\fnm{Neda} \sur{Aminnejad}}\email{neda2727@yorku.ca}
\author[1]{\fnm{Arvind} \sur{Gupta}}\email{arvind.gupta@utoronto.ca}
\author*[1,2]{\fnm{Huaxiong} \sur{Huang}}\email{hhuang@yorku.ca}

\affil*[1]{\orgdiv{Computer Science}, \orgname{University of Toronto}, \orgaddress{\street{27 King's College Cir.}, \city{Toronto}, \postcode{M5S 1A1}, \state{Ontario}, \country{Canada}}}

\affil[2]{\orgdiv{Mathematics and Statistics}, \orgname{York University}, \orgaddress{\street{4700 Keele St.}, \city{Toronto}, \postcode{M3J 1P3}, \state{Ontario}, \country{Canada}}}

\affil[3]{\orgdiv{Data Science Research Center}, \orgname{Duke Kunshan University}, \orgaddress{\street{No.8 Duke Ave.}, \city{Kunshan}, \postcode{215300}, \state{Jiangsu}, \country{China}}}

\affil[4]{\orgdiv{Mathematics, Analytics, and Data Science Lab}, \orgname{Fields Institute for Research in Mathematical Sciences}, \orgaddress{\street{222 College St.}, \city{Toronto}, \postcode{ M5T 3J1}, \state{Ontario}, \country{Canada}}}


\abstract{
This study addresses the challenge of reconstructing unseen ECG signals from PPG signals, a critical task for non-invasive cardiac monitoring. While numerous public ECG-PPG datasets are available, they lack the diversity seen in image datasets, and data collection processes often introduce noise, complicating ECG reconstruction from PPG even with advanced machine learning models. To tackle these challenges, we first introduce a novel synthetic ECG-PPG data generation technique using an ODE model to enhance training diversity. Next, we develop a novel subject-independent PPG-to-ECG reconstruction model that integrates contrastive learning, adversarial learning, and attention gating, achieving results comparable to or even surpassing existing approaches for unseen ECG reconstruction. Finally, we examine factors such as sex and age that impact reconstruction accuracy, emphasizing the importance of considering demographic diversity during model training and dataset augmentation.}

\keywords{ECG Reconstruction, Synthetic ECG-PPG Pairs, Contrastive Learning, Vector Quantization}

\maketitle
\section{Introduction} \label{sec:introduction}
The electrocardiogram (ECG) is the gold standard for cardiovascular diagnosis; however, recording ECG signals presents several challenges. Traditional ECG devices limit user mobility, and extended monitoring can cause skin irritation, require offline data processing, and demand increased user intervention \cite{tian2023}. To address these limitations, researchers have explored photoplethysmography (PPG) as a viable alternative. PPG is non-invasive, suitable for long-term, real-time monitoring, and provides insights into heart rate, heart rate variability \cite{gil2010photoplethysmography}, respiration rate \cite{johansson2003neural}, cardiac output \cite{wang2009noninvasive}, and blood pressure \cite{chua2010towards}. Consequently, PPG's role in healthcare monitoring is growing, showing promise for personal health management \cite{castaneda2018review, Tarifi2023}. Nevertheless, ECG remains the primary diagnostic standard due to its established research foundation \cite{Chiu2020}.

Given these considerations, researchers are increasingly focused on the potential for reconstructing ECG signals from PPG. The intrinsic relationship between ECG and PPG signals arises from the effect of the heart's contractions on peripheral blood volume, governed by the sinoatrial node’s electrical signals \cite{joshi2014review}. PPG's waveform characteristics and pulse intervals provide valuable cardiovascular insights \cite{castaneda2018review, Tarifi2023}. A strong correlation between PPG's peak-to-peak interval and ECG's RR interval suggests the feasibility of deriving ECG data from PPG signals \cite{castaneda2018review, Tarifi2023, Chiu2020, El-Hajj2020}. Investigating this ECG-PPG link led to new methods for ECG reconstruction from PPG, merging both technologies’ strengths and creating new possibilities in cardiovascular monitoring.

Reconstructing ECG from PPG involves estimating the ECG signal from the PPG waveform through advanced signal processing and machine learning. This task is complex due to the distinct nature of each signal, the intricate ECG-PPG relationship, and possible artifacts in ECG and PPG data. Current research explores various strategies, including time-domain, frequency-domain techniques, and deep learning. Developing efficient reconstruction algorithms holds great promise for transforming cardiovascular monitoring.

Delving deeper into existing methods, many studies have focused on deducing ECG waveforms from clean PPG signals. However, most prevalent approaches are subject-dependent \cite{Zhu2021, tian2023, Abdelgaber2023, Tang2022, Ezzat2024}, often relying on predicting future ECG cycles from the same individual rather than performing subject-independent, unseen ECG predictions. Typically, these subject-dependent methods train on a segment of ECG cycles and test on the remaining portion, allowing the model to learn specific ECG characteristics, such as waveform shape and frequency, from the individual during training.

In contrast, far fewer studies, such as \cite{Chiu2020, Sarkar2021, Lan2022PerformerAN, Shome2023, belhasin2024}, have delved deeply into subject-independent PPG-to-ECG reconstruction. \cite{Sarkar2021} introduced CardioGAN, a deep learning model based on the Generative Adversarial Network (GAN) architecture. Drawing from CycleGAN \cite{zhuCycleGAN2017}, CardioGAN uses cycle consistency loss to train without paired ECG-PPG data. However, our observations suggest that cycle consistency loss alone is insufficient, prompting us to introduce a mid-way reconstruction loss for improved results. \cite{Chiu2020} proposed an encoder-decoder framework that employs a sequence transformer to account for PPG signal variations, and an attention network to highlight critical PPG regions for ECG reconstruction. Their approach centred on a QRS complex-enhanced loss function, focuses on refining the QRS complex with a Gaussian weighting around the R peak index. Additionally, Shome et al. \cite{Shome2023} proposed the Region-Disentangled Diffusion Model (RDDM), which leverages a novel diffusion model architecture for high-fidelity PPG-to-ECG translation. The RDDM model addresses a core limitation in existing diffusion models, namely, the indiscriminate addition of noise across the entire signal, by introducing a region-specific noise process that targets critical regions of interest (ROIs), such as the QRS complex in ECG signals. This disentanglement process enables RDDM to generate high-quality ECG signals from PPG inputs in just ten diffusion steps.

Predicting subject-independent ECG signals using machine learning models poses significant challenges, partly due to the limited amount of ECG data available compared to image data. This shortage results in less diversity and a narrower population distribution in ECG datasets. In contrast, large image datasets have enabled advanced deep learning models to achieve impressive performance across various fields and real-world applications. To address ECG data limitations, creating synthetic datasets offers a viable solution, helping researchers overcome the difficulty of gathering large, diverse real-world datasets. This approach not only enhances data diversity but is also cost-effective, often proving more economical than collecting extensive real-world data. Many researchers have attempted to synthesize ECG signals using generative deep learning models, such as \cite{adib2022, Kaleli2023, Abdelgaber2023, Adib2023}. However, these models predominantly focus on generating synthetic ECG signals rather than ECG-PPG pairs. In our study, we introduce an advanced ODE-based technique to generate synthetic ECG-PPG pairs, evaluating our methods on both synthetic and real-world datasets.

In our ECG reconstruction approach, we utilize three key techniques. First, we apply Contrastive Learning to distinguish between similar and dissimilar data points within an embedded space, aligning reconstructed ECG signals from PPG data with real ECG signals. This approach enhances the model’s ability to differentiate genuine waveforms from reconstructed ones. Second, we employ adversarial learning, specifically GANs, to balance a generator that produces data resembling real samples and a discriminator that distinguishes real from generated data. This process enables the generation of realistic ECG signals from PPG inputs. Lastly, we integrate an Attention Gate (AG) model into a U-Net architecture, which highlights key regions to ensure precise extraction of ECG signals from PPG data. This combination enhances the fidelity of the reconstructed ECG.

Our deep learning framework, CLEP-GAN (an acronym for ``Contrastive Learning for ECG reconstruction from PPG signals''), integrates contrastive learning, adversarial learning, and attention gating for precise subject-independent ECG reconstruction from PPG signals. Whereas we primarily rely on the Attention U-Net for signal generation, the VQ-VAE network has also been explored. VQ-VAE merges Variational AutoEncoder (VAE) principles with Vector Quantization (VQ). Instead of directly translating the encoder's output to a continuous latent space, VQ-VAE quantizes the encoder's output to its closest codebook code, offering advantages over standard VAEs and GANs. 

The primary objective of this study is to advance subject-independent ECG reconstruction from PPG signals by addressing key limitations in current methods and datasets. Our specific goals are:
\begin{itemize}
    \item \textbf{Generate Synthetic ECG-PPG Data}: We introduce an ODE-based method for generating synthetic ECG-PPG pairs to explore ways of increasing data diversity and addressing the limitations of current ECG-PPG datasets, which often lack diversity and contain noise. Our work provides a foundation for further refinement and data augmentation strategies in ECG reconstruction research.
    \item \textbf{Develop CLEP-GAN for ECG Reconstruction}: We propose CLEP-GAN, a novel model for subject-independent ECG reconstruction from PPG signals. CLEP-GAN integrates contrastive learning, adversarial learning, attention gating, and Vector Quantized-Variational Autoencoder (VQ-VAE) components to enhance reconstruction performance and expand methodological diversity in the field.
    \item \textbf{Analyze Influential Factors in ECG Reconstruction}: We investigate the effects of factors such as sex and age on ECG reconstruction accuracy, highlighting the risks of indiscriminate dataset augmentation and underscoring the importance of carefully selected data.
\end{itemize}

\section{Dataset}
\subsection{Real dataset} \label{sec:real_dataset}
We evaluated our method's efficacy using two public real datasets. The BIDMC PPG and Respiration dataset \cite{Pimentel2017Towards, Goldberger2000PhysioBank} contains $53$ paired PPG and ECG recordings from $45$ critically ill patients at Beth Israel Deaconess Medical Centre. Each recording lasts $8$ minutes and is sampled at $125$ Hz. The dataset has data from $20$ male patients aged between $19$ and over $90$ years, with a mean age of $66$ years and a standard deviation of $17$ years.

For further validation, we employed the CapnoBase TBME RR benchmark dataset \cite{SP2/NLB8IT_2021}. It consists of $42$ eight-minute PPG and ECG recordings sampled at $300$ Hz, sourced from $29$ pediatric surgeries and $13$ adult surgeries. Each recording is associated with a distinct individual, and the PPG signal is captured from the fingertips using a pulse oximeter.

\subsection{Data processing} \label{sec:data_processing}
\begin{figure}[htbp] 
\centering
\begin{subfigure}{\textwidth}
  \centering
  \includegraphics[width=\linewidth]{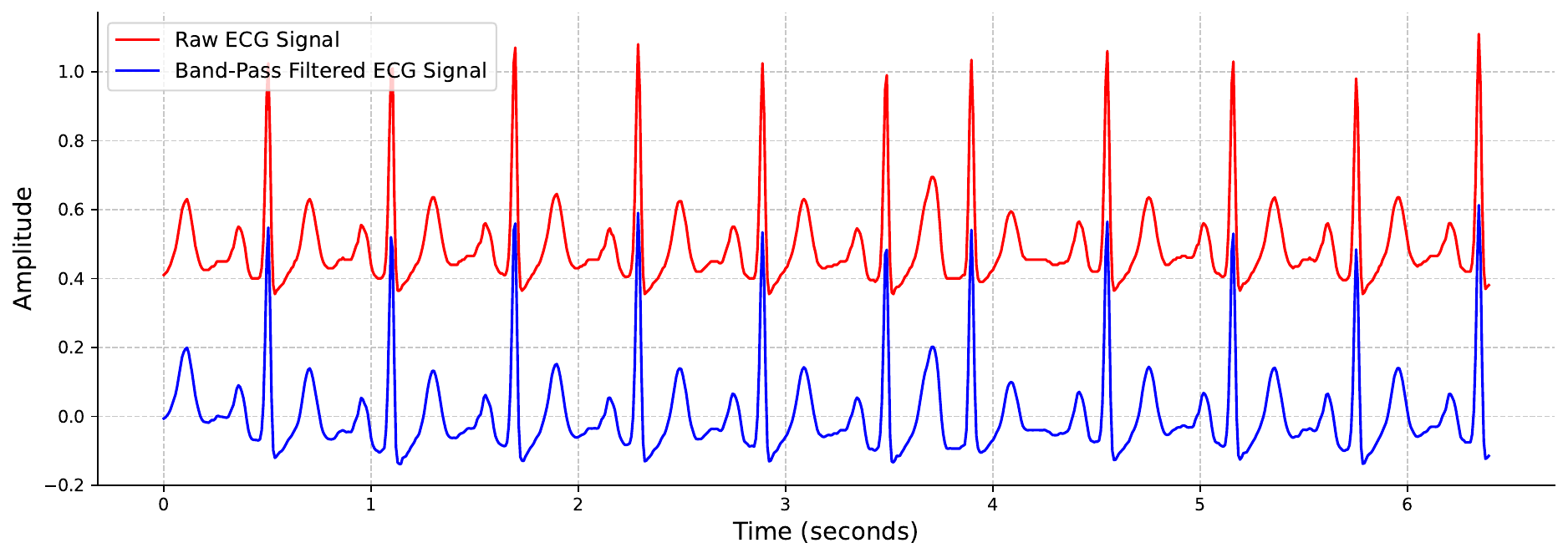}
  \caption{Raw ECG vs. filtered ECG.}
  \label{fig:preprocess_ECG}
\end{subfigure}
\begin{subfigure}{\textwidth}
  \centering
  \includegraphics[width=\linewidth]{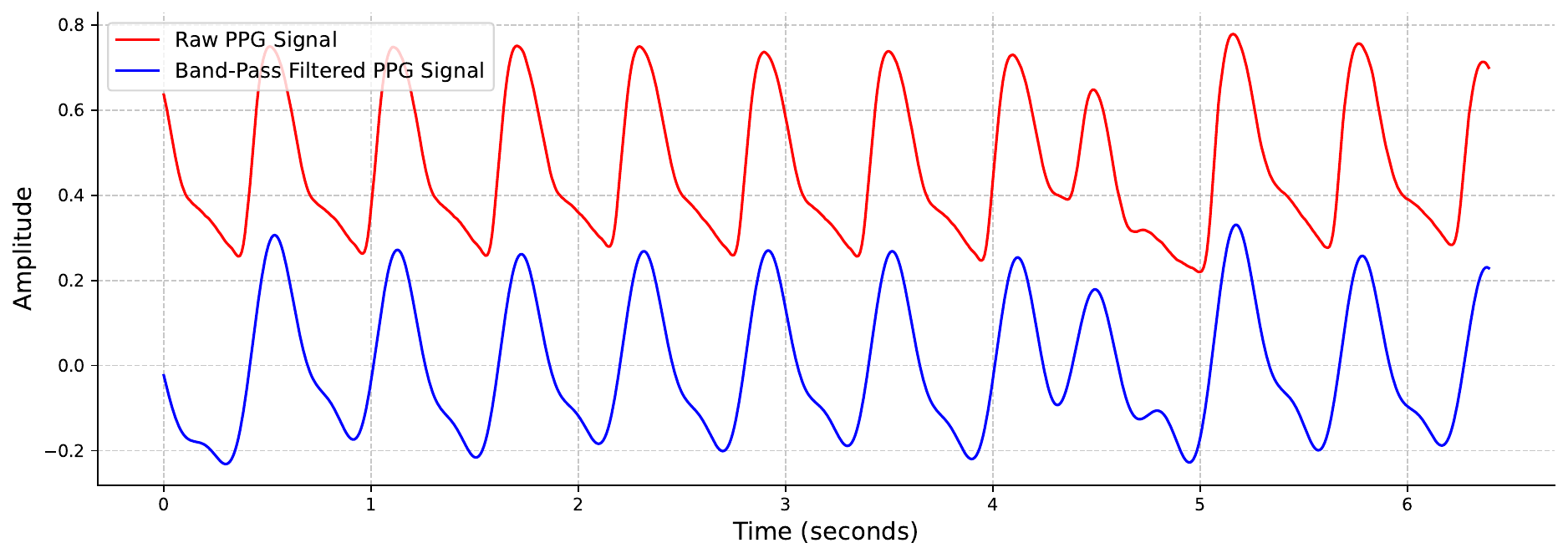}
  \caption{Raw PPG vs. filtered PPG.}
  \label{fig:preprocess_PPG}
\end{subfigure}
\caption{Comparison between raw and filtered signals. ECG signals were filtered between $0.4$ Hz and $45$ Hz, while PPG signals were filtered between $0.3$ Hz and $8$ Hz.}
\label{fig:filtered_ecg_ppg}
\end{figure}

\paragraph{Preprocessing}
To counteract noise in real datasets, we applied bandpass filters to both ECG and PPG signals. ECG signals were filtered between $0.4$ Hz and $45$ Hz, while PPG signals were filtered from $0.3$ Hz to $8$ Hz. Notably, the critical frequency range for ECG signals, containing important components like the T wave and the QRS complex, generally lies between $0.5$ Hz and $100$ Hz. The chosen range of $0.3$ Hz to $8$ Hz for PPG is appropriate, as it preserves key physiological features, such as heart rate and respiratory rate, while effectively reducing noise \cite{Shelley2007, Allen2007}. Fig. \ref{fig:filtered_ecg_ppg} illustrates the comparison between raw and filtered signals. In this example, the details of the ECG signal are fully preserved, while noise is removed from the PPG signal.

Given the different sampling rates of the two datasets, we downsampled the CapnoBase dataset from $300$ Hz to align with the BIDMC's $125$ Hz. This step ensured consistency in our analysis. Similarly, our synthetic dataset followed the sample rate of the BIDMC dataset.

\paragraph{Segmentation}
To simplify the signal pair synchronization, we aligned the initial peaks of the ECG and PPG signals, eliminating the need for intricate synchronization methods. We generated training and testing samples from our ECG-PPG dataset using a moving window, which covered $512$ data points and had a $50\%$ overlap.

For analysis uniformity, we employed min-max scalar normalization on both ECG and PPG signals, scaling their magnitudes to fall within the range of $-1$ to $1$.

\subsection{Synthetic ECG-PPG pairs} \label{sec:synthetic_data}
To generate realistic synthetic ECG signals, we make several key assumptions that enhance the fidelity and robustness of the simulated data. First, we define three primary parameters: amplitude ($\bm{a}$), width($\bm{b}$), and reference angles ($\bm{\theta}$), each parameter includes at least five elements to represent the five main ECG waves: P, Q, R, S, and T, providing a structured template that preserves typical waveform shapes. Second, we simulate the RR interval distribution to resemble that of real ECG signals by initially measuring the peak-to-peak interval distribution from the PPG signal. Assuming a consistent RR interval distribution allows the model to more accurately capture dominant R peaks, aiding in the identification of the ECG waveform structure. Finally, to approximate real-world conditions, we introduce controlled noise by varying each of the three parameters. Specifically, in addition to the main five waves, we add small, noisy waveforms by introducing extra elements in the amplitude, width, and reference angles. This controlled noise simulates common ECG artifacts, increasing the robustness of the synthetic data for training and evaluation purposes.

We utilized an Ordinary Differential Equation (ODE) model \cite{mcsharry2003dynamical}:
\begin{align}
\frac{dx}{dt} &= \alpha x - \omega y, \\
\frac{dy}{dt} &= \alpha y + \omega x, \\
\frac{dz}{dt} &= - \sum_{i \in \{ P,Q,R,S,T \}} \Big( a_i \Delta \theta_i \exp{\Big( \frac{-\Delta \theta_i^2}{2 b_i^2} \Big)} \Big) - (z - z_0), 
\end{align}
where the ECG signal is represented by $z(t)$. Here, $\alpha = 1 - \sqrt{x^2(t) + y^2(t)}$, $\theta = \text{atan2}(y(t), x(t))$, $\Delta \theta_i = (\theta - \theta_i)^2$, $\omega = 2 \pi f$, and $z_0 = A \sin(2 \pi f_0 t)$. The constants $A$ and $f_0$ are fixed at $0.01$ and $0.25$, respectively.

To correspondingly generate synthetic PPGs, we introduced two additional differential equations:
\begin{align}
\frac{dv}{dt} &= -B_0 v + B_1 w, \\
\frac{dw}{dt} &= z^2 - B_2 w.
\end{align}
The synthetic PPG signal is denoted by $v(t)$. The terms $-B_0 v$ and $-B_2 w$ induce a decay to $v(t)$ and $w(t)$, respectively. Here, $w(t)$ serves as an intermediary state variable linking the ECG signal $z(t)$ to the PPG signal $v(t)$. The values for $B_0$, $B_1$, and $B_2$ are set at $0.5$, $0.5$, and $1.25$, respectively.

In this model, three pivotal parameters: $\bm{a}$, $\bm{b}$, and $\bm{\theta}$, play key roles in simulating distinct ECG signal characteristics. These parameters characterize specific features of the P wave, Q wave, R peak, S wave, and T wave, including amplitude, width, and reference angles. Based on the ODE in \cite{mcsharry2003dynamical}, we initialized the parameters as $\bm{a} = [1.2, -5.0, 30.0, -7.5, 0.75]$, $\bm{b} = [1.2, -5.0, 30.0, -7.5, 0.75]$, and $\bm{\theta} = [-\frac{\pi}{3}, -\frac{\pi}{12}, 0, \frac{\pi}{12}, \frac{\pi}{2}]$. The initial conditions for the state vector $\bm{u}(t)=[x(t),y(t),z(t), v(t),w(t)]$ were set as $\bm{u}_0 = \left[\frac{1}{\sqrt{2}}, \frac{1}{\sqrt{2}}, 0.2, 0.005, 0\right]$. 

To simulate variations in ECG cycles, we dynamically adjusted the frequency parameter $f$ to reflect changes in RR intervals. Let $\Bar{f}$ denote the frequency corresponding to one ECG cycle associated with the average RR interval, $\overline{RR}$. When the RR interval of a specific cycle, denoted as $RR_c$, deviates from $\overline{RR}$, the frequency for that ECG cycle is recalculated as:
\begin{align}
f = \Bar{f} \times \frac{\overline{RR}}{RR_c}.
\end{align}
We initialized the value of $\bar{f}$ to $0.1$ when the $\overline{RR}$ is equal to the sampling rate, which corresponds to a heart rate of $60$ Beats Per Minute.

\subsubsection{Simulate Three Common Rhythms}
\begin{figure}[htb!] 
\centering
\begin{subfigure}[b]{0.9\textwidth}
\centering
\includegraphics[width=\linewidth]{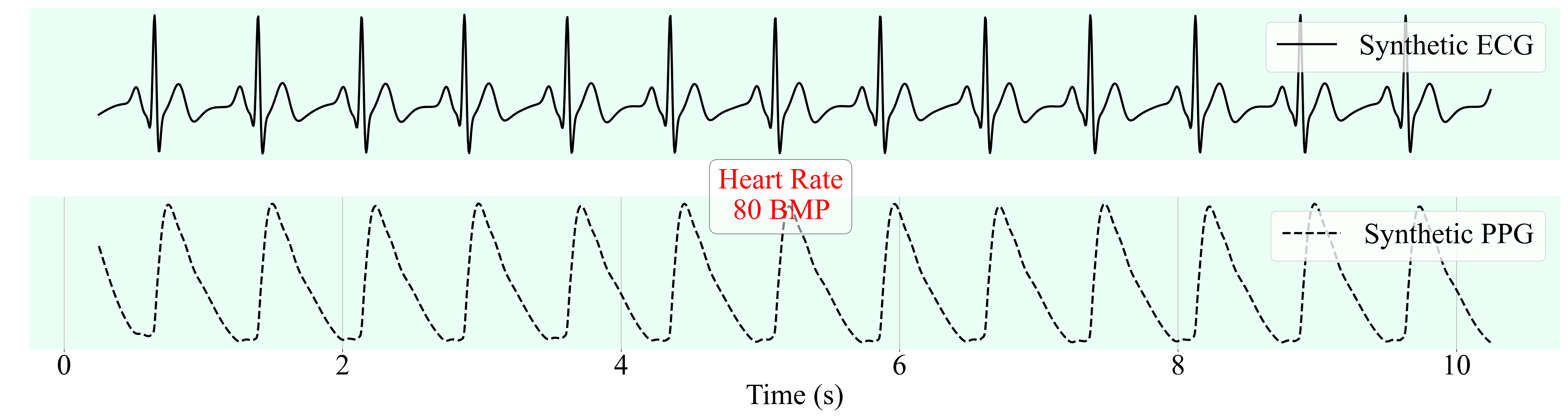}
\caption{ECG-PPG pair of RSR.}
\label{}
\end{subfigure}
\begin{subfigure}[b]{0.9\textwidth}
\centering
\includegraphics[width=\linewidth]{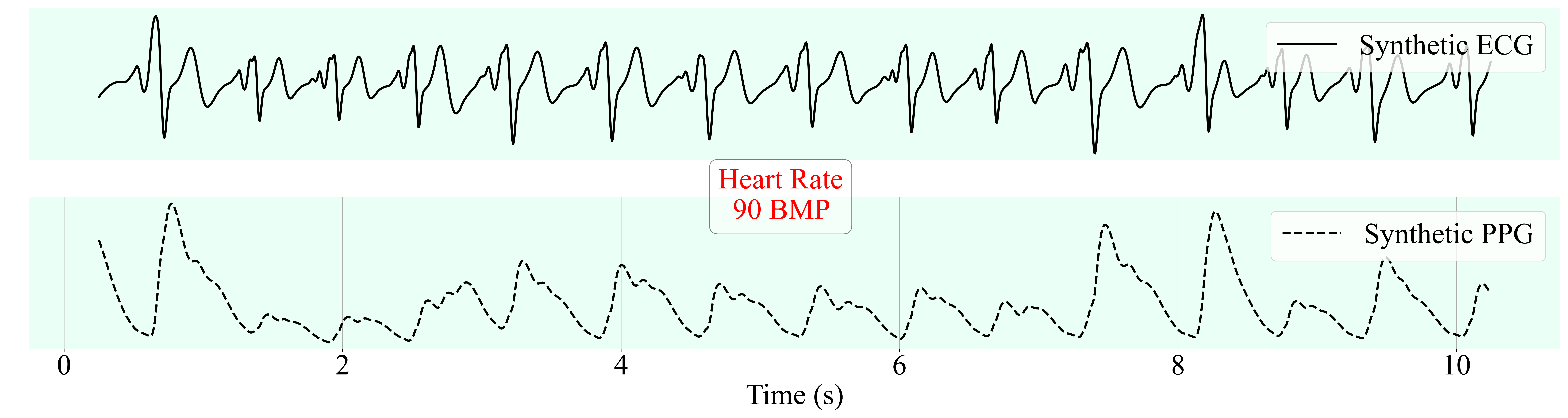}
\caption{ECG-PPG pair of SA.}
\label{}
\end{subfigure}
\begin{subfigure}[b]{0.9\textwidth}
\centering
\includegraphics[width=\linewidth]{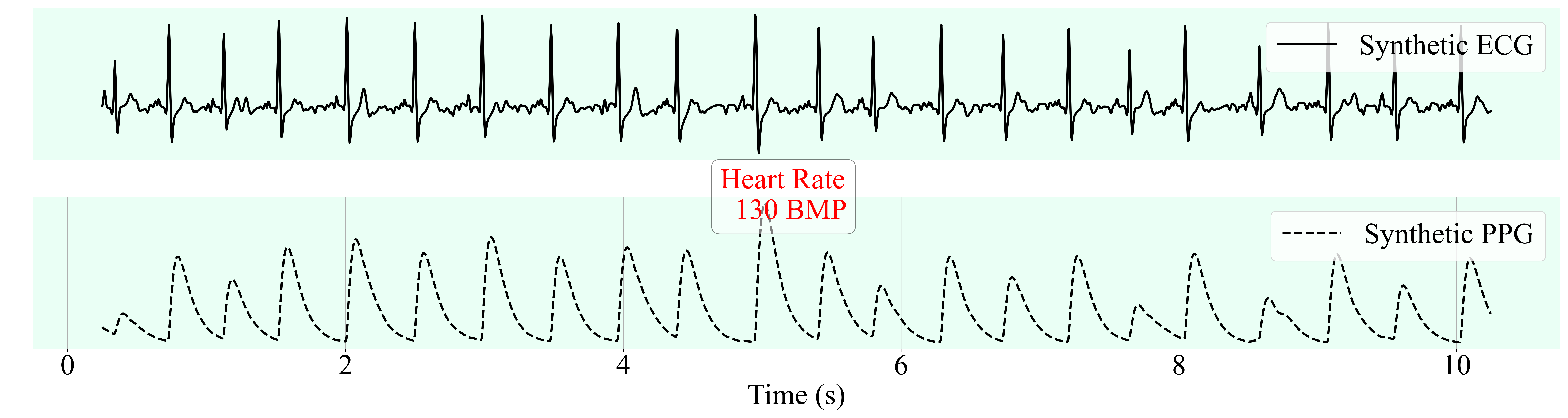}
\caption{ECG-PPG pair of AFib.}
\label{} 
\end{subfigure}
\caption{Examples of three common rhythms generated by our ODE algorithm.}
\label{fig:ecg_ppg_rhythm}
\end{figure}

Using our ODE algorithm, we produced the three dominant ECG rhythms, along with their associated PPGs. These rhythms include regular sinus rhythm (RSR), sinus arrhythmia (SA), and atrial fibrillation (AFib), as shown in Fig. \ref{fig:ecg_ppg_rhythm}. To more accurately replicate the irregular waves characteristic of SA and AFib rhythms, we introduced extra variables for each waveform parameter.

\paragraph{Regular Sinus Rhythm (RSR)}
The RSR, primarily seen in adults, has heart rates between $60$ to $100$ beats per minute (BPM). The QRS complex in RSR is typically narrow and is accompanied by upright P waves in Lead II. In the RSR waveform generation, the parameters $\bm{a}$, $\bm{b}$, and $\bm{\theta}$ retain their initial values.

\paragraph{Sinus Arrhythmia (SA)}
SA, a harmless rhythm, is frequently observed in children. Its heart rate is akin to the standard resting rate. The rhythm usually has a narrow QRS complex with upright P waves in Lead II. For replicating the erratic P waves typical of SA, we added two variables for every waveform parameter. The parameter values are:
\begin{align*}
    \bm{a} & = [1.0, 2.0, 3.0, 3.0, 2.5, -1.0, 0.5], \\
    \bm{b} & = [0.2, 0.15, 0.15, 0.2, 0.15, 0.2, 0.4], \\
    \bm{\theta} & = [-\frac{\pi}{1.5}, -\frac{\pi}{2.0}, -\frac{\pi}{6.5}, -\frac{\pi}{12.0},  0,  \frac{\pi}{12.0}, \frac{\pi}{1.5}].
\end{align*}

\paragraph{Atrial Fibrillation (AFib)}
AFib. is an erratic heart rhythm marked by unstructured QRS complexes. Its distinguishing features include the chaotic rhythm and the lack of P waves. The heart rate in AFib. varies widely due to individual factors and specific situations. To emulate the erratic nature of P and T waves in AFib., we added six additional variables for every waveform parameter. The parameter values are:
\begin{align*}
\bm{a} & = [-1.0, 0.5, 1.0, -2.0, 25.0, -10.0, 2.0, -2.0, 0.5, 0.5, 0.5],  \\
\bm{b} & = [0.1, 0.15, 0.1, 0.1, 0.1, 0.1, 0.1, 0.1, 0.2, 0.2, 0.2], \\
\bm{\theta} & = [-\frac{\pi}{2.0}, -\frac{\pi}{3.0}, -\frac{\pi}{5.0}, -\frac{\pi}{12.0}, 0, \frac{\pi}{12.0}, \frac{\pi}{6.0}, \frac{\pi}{5.0}, \frac{\pi}{2.5}, \frac{\pi}{2.0}, \\
&\frac{\pi}{1.5}].
\end{align*}

\subsubsection{Simulate RR Interval Distribution} \label{sec:simulate_RR_distribution}
From the available ECG-PPG pairs in the BIDMC dataset, we carefully selected $34$ pairs that exhibit minimal noise interference. For each rhythm (SRS, SA, and AF), we have 34 unique ECG-PPG pairs, each distinguished by distinct heart rates and heart rate variabilities. To infuse variation into the waveforms of each signal pair, we added white noise to the three parameters: \( \bm{a} \), \( \bm{b} \), and \( \bm{\theta} \). This noise has a mean of 0 and a standard deviation equal to 10\% of the original parameter values. This noise introduction ensures the synthesized signals' diversity and realism.

\begin{figure}[htbp] 
\centering
\begin{subfigure}{0.7\textwidth}
  \centering
  \includegraphics[width=\linewidth]{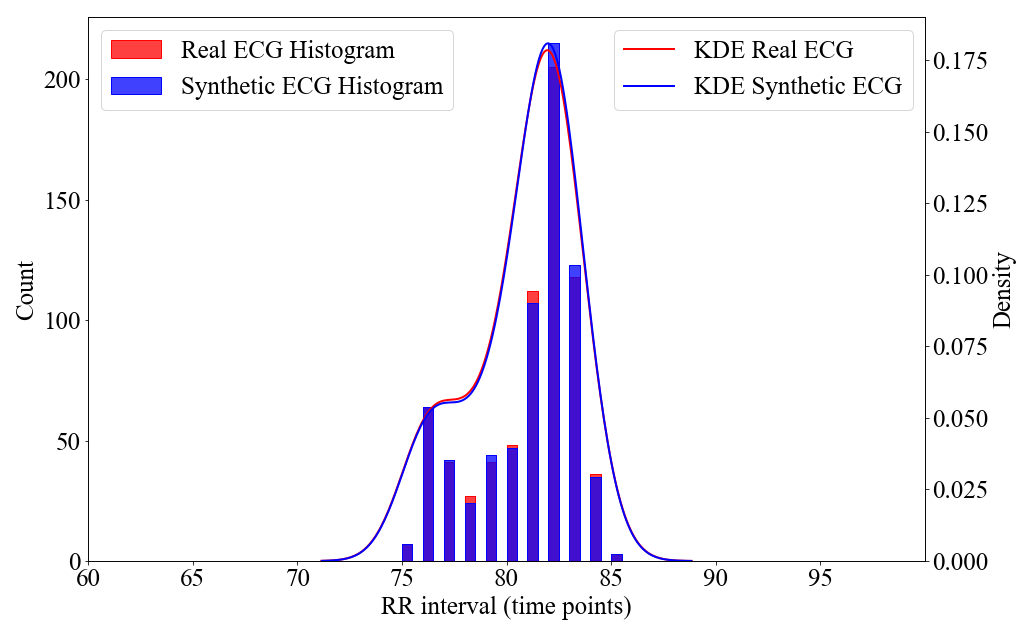}
  \caption{Record 12}
  \label{fig:RR_distribution_12}
\end{subfigure}
\begin{subfigure}{0.7\textwidth}
  \centering
  \includegraphics[width=\linewidth]{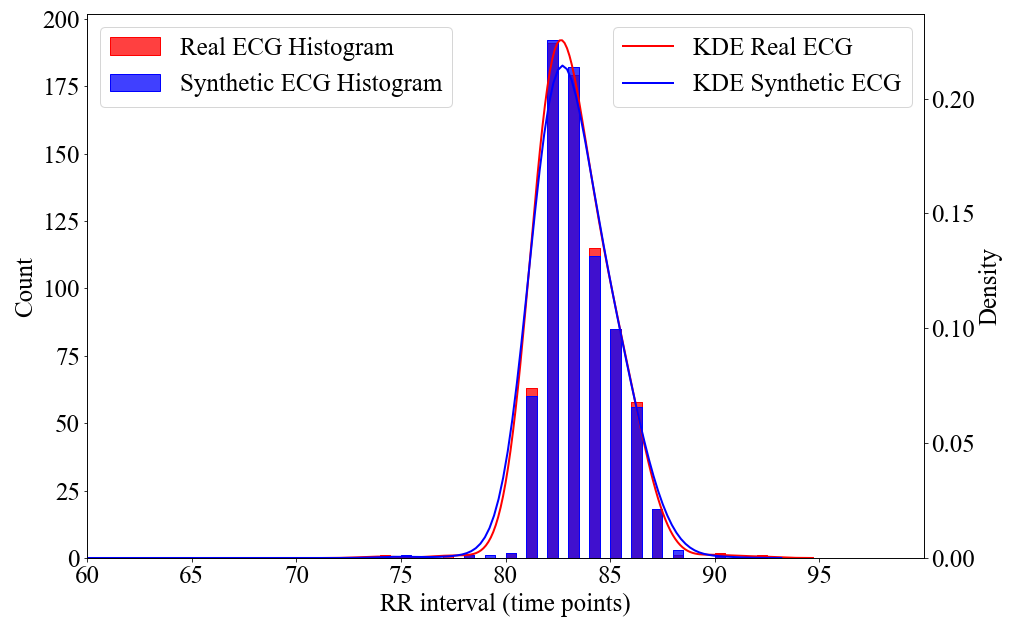}
  \caption{Record 51}
  \label{fig:RR_distribution_51}
\end{subfigure}
\caption{Comparison of RR interval distributions between synthetic ECGs and actual ECGs. For peak detection, we utilized the peak finding algorithm from NeuroKit2 \cite{Makowski2021neurokit}. The real ECG signals originate from two different records: record 12 and record 51, within the BIDMC dataset.}
\label{fig:RR_interval_distribution}
\end{figure}

\begin{figure*}[htbp] 
\centering
\begin{subfigure}{0.7\textwidth}
  \centering
    \includegraphics[width=\linewidth]{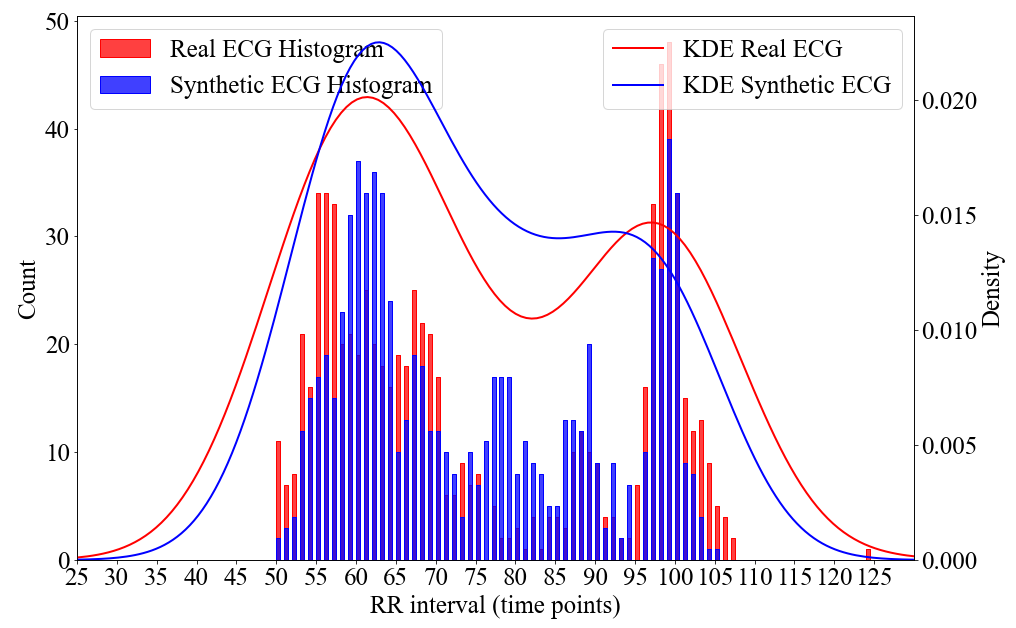}
    \includegraphics[width=\linewidth]{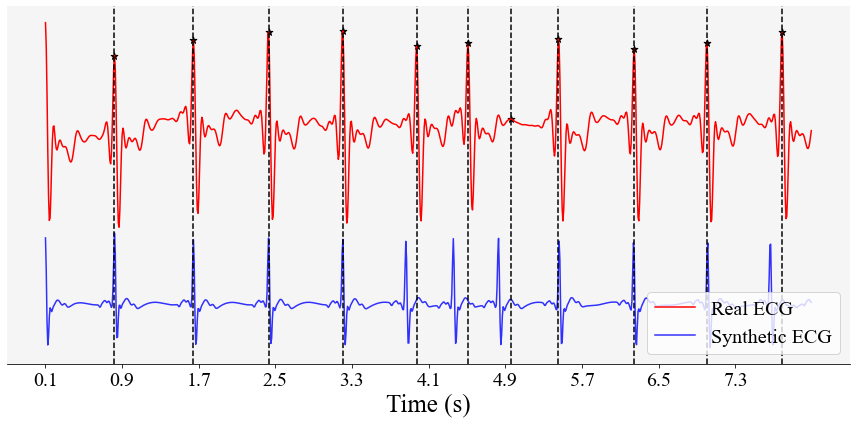}
  \caption{Record 17}
  \label{fig:ECG_real_synthetic_17}
\end{subfigure}
\begin{subfigure}{0.7\textwidth}
  \centering
 \includegraphics[width=\linewidth]{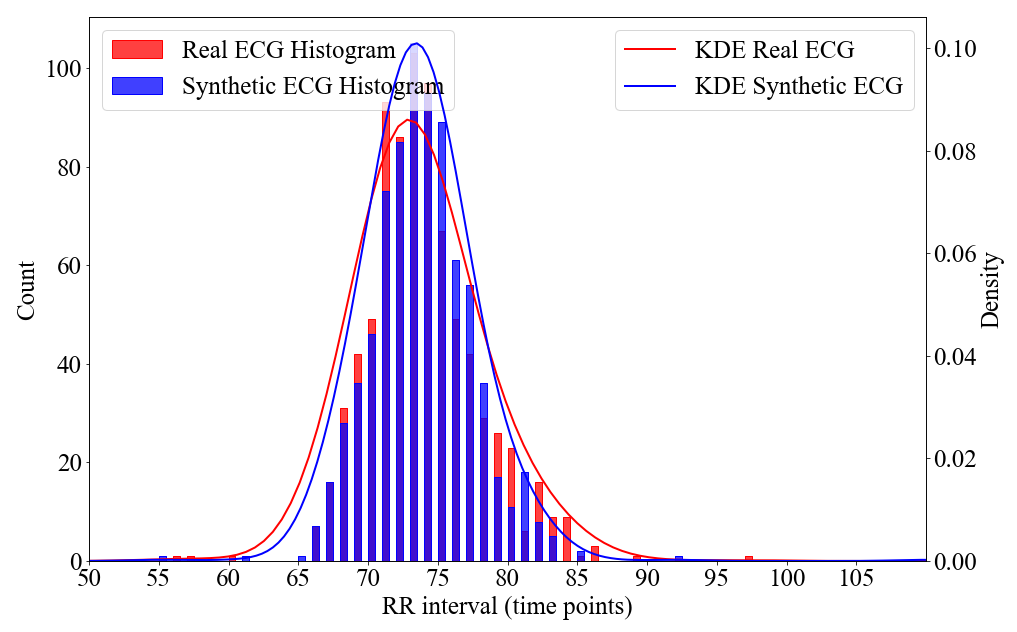}
 \includegraphics[width=\linewidth]{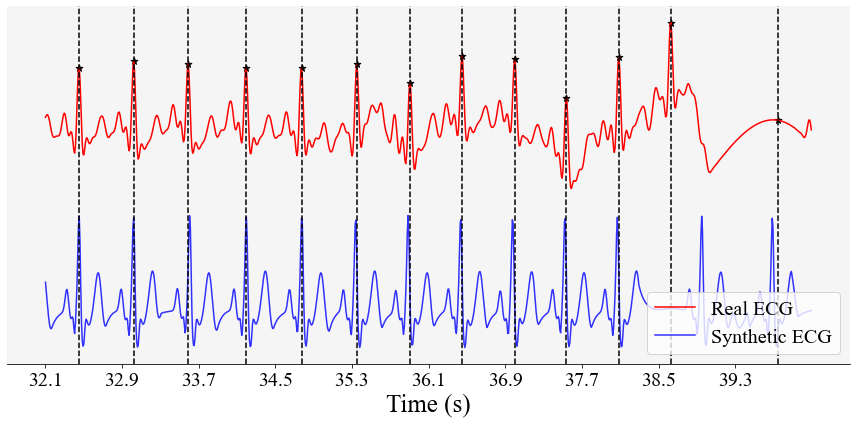}
  \caption{Record 35}
  \label{fig:RR_distribution_35}
\end{subfigure}
\caption{Illustration of RR interval discrepancies between two pairs of synthetic and real ECG signals. The real signals are derived from the BIDMC dataset. Both the black star markers and the vertical black dashed lines indicate the locations of R peaks in the real ECG. For peak detection, we employed the peak finding algorithm from NeuroKit2.}
\label{fig:RR_interval_discrepency}
\end{figure*}

In Fig. \ref{fig:RR_interval_distribution}, we present a comparison of the RR interval distributions between synthetic and real ECGs. The histograms and Kernel Density Estimation (KDE) lines for the real and synthetic RR intervals overlap significantly, suggesting a high degree of similarity. Fig. \ref{fig:RR_interval_discrepency} illustrates the discrepancies in RR intervals between two pairs of synthetic and real ECG signals. For each subfigure, the lower portion showcases a segment of the ground-truth (or real) ECG aligned with its corresponding synthetic ECG, designed to emulate its RR interval distribution. The locations of R peaks in the real ECG are denoted by both black star markers and vertical black dashed lines.

From these ECG segments, a notable observation emerges: a primary source of significant discrepancies in RR intervals between real and synthetic ECGs stems from the current constraints of the peak finding algorithm. Specifically, when faced with ECG noise (as depicted in Fig. \ref{fig:RR_distribution_35}) or the irregular RR interval patterns characteristic of Atrial Fibrillation ECGs (as seen in Fig. \ref{fig:ECG_real_synthetic_17}), the peak-finding algorithm occasionally introduces additional false ``R peaks''. This unexpected insertion subsequently prompts the synthetic algorithm to alter its frequency in an attempt to accommodate these erroneous R peaks. Therefore, the efficacy of this method is closely tied to the accuracy and inherent limitations of the peak-finding algorithm. Any shortcomings within the peak-finding methodology might affect the overall fidelity of the simulation.


\subsubsection{Quantitative Evaluation} \label{sec:synthetic_rr_evaluation}
We employed five metrics to assess the similarity between the distributions of actual RR intervals and their synthetic counterparts: relative Histogram Intersection (rHI), relative Root Mean Square Error (rRMSE), relative Earth Mover's Distance (rEMD), Kullback-Leibler Divergence (KL), and the Two-sample Kolmogorov-Smirnov (KS) test. Given the impact of bin size on metric outcomes, we standardized bin widths to a unit distance.

\paragraph{Relative Histogram Intersection (rHI)}

The Histogram Intersection (HI) between histograms \(A\) and \(B\) is given by \(\mathrm{HI}(A,B) = \sum_{i} \min(\text{count}_A(i), \text{count}_B(i))\), where \(\text{count}_A(i)\) and \(\text{count}_B(i)\) are counts in the \(i\)th bin. The relative Histogram Intersection (rHI) is 
\begin{align}
    \mathrm{rHI} = \frac{\mathrm{HI}(A, B)}{\min(\sum A, \sum B)},
\end{align}
which produce values in \([0, 1]\) with $1$ indicating identical histograms.

\paragraph{Relative Root Mean Square Error (rRMSE)}

The Root Mean Square Error (RMSE) between distributions \(A\) and \(B\) with \(N\) real RR intervals \(A_i\) is given by: $\mathrm{RMSE(A, B)} = \sqrt{\frac{1}{N}\sum_{i=1}^{N} (A_{i} - B_{i})^2}.$ Using the mean of the observed intervals \(\mu_{A} = \frac{1}{N}\sum_{i=1}^{N} A_{i}\), the normalized RMSE (rRMSE) is:
\begin{align}
\mathrm{rRMSE} = \frac{\mathrm{RMSE(A, B)}}{\mu_{A}}.
\end{align}

\paragraph{Relative Earth Mover's Distance (rEMD)}

The Earth Mover's Distance (EMD) quantifies the effort to transform one distribution into another and is derived from the cumulative distribution functions \( F_A \) and \( F_B \) of distributions \( A \) and \( B \) as:
\begin{align}
\mathrm{EMD}(A, B) = \int_{-\infty}^{\infty} |F_A(x) - F_B(x)| dx.
\end{align}

The Relative Earth Mover's Distance (rEMD) normalizes the EMD by the maximum possible EMD (MaxEMD), given by \(\mathrm{TotalEarth} \times \mathrm{MaxDistance}\), where \(\text{TotalEarth}\) is one histogram's total count and \(\text{MaxDistance}\) is the largest distance between bins:
\begin{align}
\mathrm{rEMD} &= \frac{\mathrm{EMD}(A, B)}{\mathrm{MaxEMD}}.
\end{align}
A smaller rEMD suggests greater similarity between the distributions.

\paragraph{Kullback-Leibler Divergence (KL)}
KL divergence measures the difference between two probability distributions \(P\) and \(Q\). For discrete distributions, it's computed as:
\begin{align}
KL(P || Q) = \sum_{i} P(i) \log \left( \frac{P(i)}{Q(i)} \right).
\end{align}
It's noteworthy that KL divergence is asymmetric.

\paragraph{Kolmogorov-Smirnov (KS) Test}
The KS test quantifies the largest difference between the cumulative distribution functions (CDFs) of two samples:
\begin{align}
D_n = \sup_x |F_1(x) - F_2(x)|,
\end{align}
with \(F_1(x)\) and \(F_2(x)\) being the empirical distribution functions of the two samples.

Our evaluation, as outlined in Table \ref{tab:RR_distribution_evaluation}, utilizes three peak-finding algorithms: Neurokit \cite{Makowski2021neurokit}, Scipy's \texttt{signal.find\_peaks} with a ``distance'' parameter set to $50$, and the Hamilton segmenter \cite{hamilton2002opensource} from BioSPPy \cite{carreiras2015biosppy}. Prior to peak detection with Neurokit, the raw ECG signals undergo a cleaning process using NeuroKit2's ECG clean method. Both Neurokit and the Hamilton segmenter are applied with default settings. The results in Table \ref{tab:RR_distribution_evaluation} indicate a close alignment between the synthetic and ground truth RR interval distributions. Whereas our chosen algorithms produce consistent results, a more refined peak-finding method could further improve the fidelity of RR interval simulations.

\begin{table}[htb!] 
\small
\centering
\setlength{\tabcolsep}{4pt}
\begin{tabular}{@{}ccccccc@{}} 
 \toprule
  Methods & Data & rHI $\uparrow$  & rRMSE $\downarrow$ & rEMD $\downarrow$ & KL $\downarrow$ & KS $\downarrow$\\
 \midrule 
 & Data 1 &  0.95 & 0.05 & 7.05e-5 &  0.25 & 0.02\\ 
  \multirow{-2}{*}{Neurokit}  & Data 2 & 0.97 & 0.03  & 7.76e-5 &  0.10 & 0.01\\
 \midrule
  SciPy  & Data 1 & 0.95 & 0.04  & 5.19e-5  & 0.19  & 0.02\\ 
  Peak Finding & Data 2 & 0.97 & 0.03  & 5.04e-5  & 0.12  & 0.01\\     
 \midrule
   Hamilton & Data 1 & 0.95 & 0.05  & 6.92e-5  & 0.28  & 0.02\\ 
   Segmenter & Data 2 & 0.97 & 0.03 & 7.38e-5  &  0.11 & 0.01\\ 
 \bottomrule
\end{tabular}
\vspace{1em}
\caption{We evaluated the discrepancy in the RR interval distribution between synthetic and real ECGs using three peak-finding algorithms: the Neurokit algorithm, the Scipy peak-finding algorithm, and the Hamilton segmenter. Discrepancies were measured using five metrics: relative Histogram Intersection (rHI), relative Root Mean Square Error (rRMSE), relative Earth Mover's Distance (rEMD), Kullback-Leibler Divergence (KL), and the Two-sample Kolmogorov-Smirnov test (KS). ``Dataset 1'' consists of all $34$ ECGs from the BIDMC Dataset, while ``Dataset 2'' contains $30$ ECGs, excluding the four signals having a challenge of peak-finding as shown in Fig. \ref{fig:RR_interval_discrepency}. Results represent the average for each dataset.}
\label{tab:RR_distribution_evaluation}
\end{table}

\section{Method} \label{sec:method}
\subsection{Proposed Architecture} \label{sec:method_architecture}
\begin{figure}[htb!] 
\centering
\includegraphics[width=\linewidth]{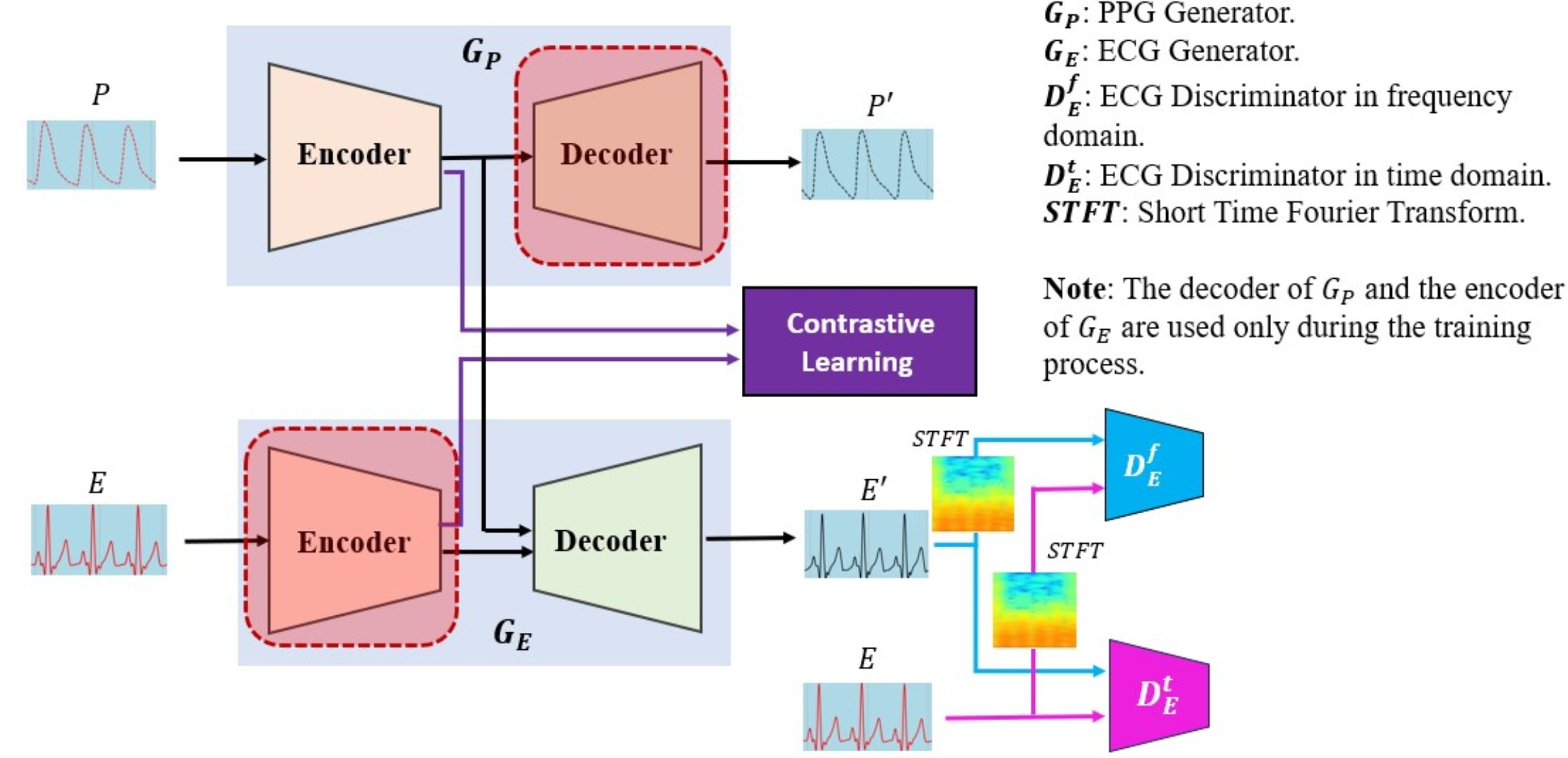}
\caption{The architecture of our method.}
\label{fig:architecture}
\end{figure}

Fig. \ref{fig:architecture} illustrates the architecture of our proposed method, which comprises four primary components: ECG generator ($G_E$), PPG generator ($G_P$), time-domain-based discriminator ($D_E^t$), and frequency-domain-based discriminator ($D_E^f$). Contrary to traditional strategies that directly convert PPG to ECG, this work leverages contrastive learning. Specifically, the PPG generator ($G_P$) reconstructs a generated PPG ($P'$) from the actual PPG ($P$), while the ECG generator ($G_E$) produces a generated ECG ($E'$) from the real ECG ($E$).

During training, both $G_E$ (the ECG generator) and $G_P$ (the PPG generator) are used to reconstruct the PPG and ECG signals, respectively. After training, only the encoder of $G_P$ and the decoder of $G_E$ are utilized for inference. The contrastive learning component is optimized during training to bring similar (positive) features closer together, aligning the encoder output of $G_E$ with that of $G_P$, while maintaining separation between dissimilar (negative) features. In this ECG and PPG scenario, positive pairs represent the same cardiac event across both signal types, encouraging these pairs to be closer in the latent space, while negative pairs, mismatched ECG and PPG representations, are kept apart. The decoder of $G_E$ is provided with two inputs to calculate two distinct reconstruction losses: the ECG-to-ECG loss and the ECG-to-PPG loss.

Given the notable success of GANs in generation tasks, we employ GAN-based adversarial learning with the discriminator bolstering the reconstruction. Recognizing the significance of cardiac activity in both time and frequency domains, as emphasized in \cite{penttila2001time} and \cite{Sarkar2021}, we deploy two discriminators: $D_E^t$ (time domain) and $D_E^f$ (frequency domain). This approach effectively captures cardiac nuances, enhancing the accuracy of reconstructions. The detailed architectures can be found in Appendix, as illustrated in Fig. \ref{fig:Discriminators}. For frequency-domain considerations, the Short-Time Fourier Transformation (STFT) is applied to the ECG and PPG data.

In our framework, we use the Attention U-Net \cite{AttentionUNet} as the generator component, which is denoted as $G_E$ or $G_P$ (refer to Fig. \ref{fig:architecture}). The U-Net architecture has undergone various enhancements since its original introduction \cite{U-Net}. Several improved versions, such as U-Net++ \cite{U-Net++}, R2U-Net \cite{Mubashar2022}, Attention U-Net, ResUnet \cite{ResUNet}, TransUNET \cite{TransUNet}, and Swin-UNET \cite{cao2021swinunet}, have substantially outperformed the original U-Net model. We chose the Attention U-Net due to its relatively streamlined design among the advanced U-Net variants, a decision also supported by \cite{Sarkar2021}. A distinctive feature of the Attention U-Net is its attention gates, which act like a ``spotlight''. These gates highlight the most critical parts of the input data and dim the less relevant areas. This ``spotlight'' feature is invaluable for PPG-ECG reconstructions, ensuring that the model focuses intently on the nuanced ECG patterns within the PPG data. A detailed illustration of the Attention U-Net structure is available in the Appendix, as shown in Fig. \ref{fig:Attention_UNet}

We propose an alternative generator architecture, denoted as \( G_E \) or \( G_P \), based on the VQ-VAE framework as described in \cite{VQVAE2017}. VQ-VAE is a variant of the autoencoder that excels at deriving compact data encodings, which are crucial for reducing dimensionality and enhancing the representation of ECG data \cite{Vincent2008, NIPS2012_6cdd60ea, Lin2023, Chen2024}. This architecture combines Vector Quantization (VQ) with the core principles of an autoencoder. By employing VQ, continuous latent variables are mapped to a discrete set, creating a more organized and consistent latent space. This discrete mapping is especially valuable for ECG signals, where even subtle variations can be clinically meaningful, as it ensures that key patterns are effectively captured and highlighted. Further details on the VQ-based discrete latent representation and the VQ-VAE architecture can be found in the Appendix, as illustrated in section \ref{sec:discrete_latent_variables} and Fig. \ref{fig:VQVAE}.

To distinctly label our two proposed architectures, we have named the one based on Attention U-Net as CLEP-GAN, and the one based on VQ-VAE as CLEP-VQGAN. ``CLEP'' stands for ``Contrastive Learning for ECG reconstruction from PPG signals''.

\subsection{Objective} \label{sec:loss}
Taking advantage of the end-to-end training process within our framework, our ultimate optimization objective is to minimize a composite loss that encompasses generation loss (comprising contrastive and adversarial losses) and reconstruction loss.
\subsubsection{Generation Loss}
For the ECG reconstruction, we employ both the contrastive loss and the reconstruction loss. Initially, we utilize the NT-Xent loss (Normalized Temperature-scaled Cross-entropy Loss) to serve as the contrastive loss from PPG to ECG. This is expressed as:
\begin{align}
L_{\text{contrast}}(z_p, z_e) = -\log \left( \frac{\exp\left(\frac{z_p \cdot z_e}{\|z_p\|_2 \times \|z_e\|_2} \right)/\tau}{\sum_{k=1}^{N} \exp\left(\frac{z_p \cdot z_k}{\|z_p\|_2 \times \|z_k\|_2} \right)/\tau} \right)
\end{align}
Here, $z_p$  denotes the latent feature representation of the PPG, while $z_e$ symbolizes the latent feature representation of the ECG. The denominator comprises the sum of exponential similarity scores between the PPG $z_p$ and all ECG representations in the batch. $N$ represents the total number of samples in the batch, and $\tau$ is a temperature parameter that controls the scale of the similarity scores. Based on empirical assessment, we selected a value of $0.1$ for $\tau$ in our experiments.

In our approach, we use three distinct reconstruction losses. The first, $L_{p2p}$, measures the discrepancy in the model’s replication of the original PPG signal, where $P' = G_P(P)$ represents the reconstructed signal based on the ground-truth PPG. The second, $L_{e2e}$, quantifies the loss associated with replicating the ECG signal from its ground-truth version, with $E' = G_E(E)$ representing the reconstructed ECG output. Finally, $L_{p2e}$ represents the loss in reconstructing the ECG from the PPG, expressed as $E' = G_E(Z_p)$, where $Z_p = EN_{G_P}(P)$, with $EN_{G_P}$ denoting the encoder component of the PPG generator network.
\begin{align}
L_{p2p} &= \mathbb{E}_{p\sim P}[|| G_P(p) - p||_{\text{smooth\_{$L_1$}}}], \\
L_{e2e} &= \mathbb{E}_{e\sim E}[|| G_E(e) - e||_{\text{smooth\_{$L_1$}}}], \\
L_{p2e} &= \mathbb{E}_{z_p\sim Z_p}[|| G_E(z_p) - e||_{\text{smooth\_{$L_1$}}}].
\end{align}
In the equations above, the symbol $\mathbb{E}_{z_p\sim Z_p}$ calculates the ``average'' reconstruction loss when using values $z_p$ from the distribution $Z_p$. Given these individual losses, the overall generation loss, $L_{\text{gen}}$, is the summation of the contrastive loss and the three reconstruction losses: $L_{\text{gen}} = L_{\text{contrast}} + L_{p2p} + L_{e2e} + L_{p2e}$. 

\paragraph{VQ loss}
The loss function for the VQ-VAE consists of three key components: reconstruction loss (or data term), dictionary loss, and commitment loss. 

\begin{enumerate}
    \item \textbf{Reconstruction Loss}: This loss is pivotal in the optimization process for both the decoder and the encoder. Due to the straight-through gradient estimation associated with the mapping from \(z_e(x)\) to \(z_q(x)\), it's evident that the embeddings, denoted by \(e_i\), aren't influenced by the gradients from the reconstruction loss. That is, during the backward pass the gradient, $\nabla_{z}L$, is transmitted unaltered back to the encoder.
    
    \item \textbf{Dictionary Loss}: To promote the learning of the embedding space, we resort to one of the most fundamental dictionary learning algorithms---Vector Quantisation (VQ). This objective, termed as the dictionary loss, leverages the \(l_2\) error to align the embedding vectors \(e_i\) with the encoder outputs \(z_e(x)\).

    \item \textbf{Commitment Loss}: A significant consideration is the unbounded nature of the embedding space, which can expand indefinitely if the embeddings \(e_i\) do not adapt alongside the encoder parameters. To mitigate this, we need a commitment loss. This ensures that the encoder remains committed to an embedding and limits its output expansion.
\end{enumerate}

Thus, the aggregate training objective is represented by:
\begin{equation}
L = \log p(x|z_q(x)) + \| \bm{sg}[z_e(x)] - e \|_2^2 + \lambda \| z_e(x) - \bm{sg}[e] \|_2^2,
\end{equation}
Here, \(\bm{sg}\) is the stop-gradient operator. It operates as an identity during the forward computation but has zero partial derivatives, effectively treating its operand as a constant that cannot be updated. In our experiments, Based on \cite{VQVAE2017}, the algorithm is robust to variations in \(\lambda\), with minimal changes in results when \(\lambda\) is adjusted between $0.1$ and $2.0$. For our work, we settled on a \(\lambda\) value of $0.25$, in line with recommendations from the original paper.

When utilizing VQ-VAE as our generator within the framework, it's essential to note that the reconstruction loss is already captured by the term \(\log p(x|z_q(x))\). Therefore, to compute the generation loss for the VQ-VAE generator, we simply need to integrate the dictionary and commitment losses. This results in the following equation for the generation loss:
\begin{align}
L_{\text{gen}} = L_{\text{contrast}} &+ L_{p2p} + L_{e2e} + L_{p2e} + \| \bm{sg}[z_e(x)] - e \|_2^2 \nonumber \\
& + \lambda \| z_e(x) - \bm{sg}[e] \|_2^2,
\end{align}

\subsubsection{Adversarial Loss}
Reconstructing PPGs and ECGs from their respective ground truths, i.e., $P' = G_P(P)$ and $E' = G_E(E)$, is a comparatively straightforward task, therefore, we do not use adversarial learning for these reconstructions. However, for the more challenging task of PPG-to-ECG reconstruction, we employ adversarial learning to improve authenticity and quality. As discussed in section \ref{sec:method_architecture}, we leverage dual discriminators for this purpose: $D_E^t$ for the time domain and $D_E^f$ for the frequency domain.

The adversarial losses associated with these discriminators are defined as follows:

For the time-domain discriminator, \(D_E^t\),
\begin{align}
L_{\text{t}} &= \mathbb{E}_{e \sim E}[\log (D^{t}_E(e))] \nonumber \\
&\quad + \mathbb{E}_{p \sim P}[\log (1 - D^{t}_E(G_E(z_p)))]
\end{align}

For the frequency-domain discriminator, \(D_E^f\), where \(STFT_{spect}(.)\) represents the Short-Time Fourier Transform, capturing spectral content,
\begin{align}
L_{\text{f}} &= \mathbb{E}_{e \sim E}[\log (D^{f}_E(STFT_{spect}(e)))] \nonumber \\
&\quad + \mathbb{E}_{p \sim P}[\log (1 - D^{f}_E(STFT_{spect}(G_E(z_p))))] 
\end{align}
Here, the symbol $\mathbb{E}_{p\sim P}$ calculates the ``average'' reconstruction loss when using values $p$ from the distribution $P$.

\subsubsection{Composite Loss Function}
The composite loss function aggregates the generation loss and the adversarial losses from both time and frequency domains. The composite loss function is mathematically expressed as:
\begin{align}
    L_{total} = \alpha L_{\text{gen}} + \beta L_{\text{t}} + \gamma L_{f},
\end{align}
To achieve a balanced optimization process, we use coefficients $\alpha$, $\beta$, and $\gamma$ to determine the relative importance of each loss term within the composite loss function. These coefficients can be adjusted based on the characteristics of a specific dataset to ensure optimal performance. In our study, after empirical evaluation, we have chosen the values $\alpha = 30$, $\beta = 3$, and $\gamma = 1$. This choice aligns with the findings of \cite{Sarkar2021}.

\section{Experiments} \label{sec:experiments}
\subsection{Evaluation Metrics}
In addition to the five metrics (rHI, rRMSE, rEMD, KL divergence, and KS test) mentioned in Section \ref{sec:synthetic_rr_evaluation}, our experiments employ the following metrics to assess the performance of our methods.

\paragraph{Root Mean Square Error (RMSE)}
Contrary to the rRMSE, which measures the RR interval distribution, the RMSE in this context quantifies the disparity between the predicted and actual signals. It's formally defined by:
\begin{align}
RMSE = \sqrt{\frac{\sum^n_{i=1}(S_G(i) - S_R(i))^2}{n}},
\end{align}
where $S_G$ denotes the ground-truth signal and $S_R$ the reconstructed signal. $n$ represents the length of the signals.

\paragraph{Heart Rate Variability (HRV)}
HRV measures the variability in time between consecutive heartbeats. Commonly, the mean of RR intervals and their standard deviation (STD) are employed for HRV analysis.

\paragraph{Mean Absolute Error (MAE) for Heart Rate (HR)}
The heart rate is calculated as the inverse of the RR interval, converted into beats per minute (BPM), expressed as $HR (BMP) = 60 / RR (seconds)$.

To evaluate the accuracy of heart rate relative to a ground-truth HR, we use the mean absolute error (MAE) metric. MAE quantifies the difference in heart rate derived from an ECG or PPG signal relative to the ground-truth HR. Its formula is:

\begin{align}
MAE_{HR} = \frac{1}{N}\sum_{i=1}^N |HR_i^{G}-HR_i^{R}|.
\end{align}

In this equation, $N$ indicates the total number of RR intervals from which HR measurements are obtained. The index $i$ refers to each specific interval. Meanwhile, $HR_i^{G}$ and $HR_i^{R}$ correspond to the ground truth and the reconstructed heart rates, respectively.

\paragraph*{Fr\'{e}chet Distance (FD)}
To evaluate the similarity between the generated and real ECG signals, we calculate the Fr\'{e}chet Distance (FD) \cite{alt1995computing, Sarkar2021} in feature space. The FD metric measures the statistical distance between the feature distributions of the real and generated ECG signals, taking into account both the mean and covariance of these distributions.

Given two sets of feature representations extracted from real and generated ECG signals, the FD is defined as:
\begin{align}
FD = \| \bm{\mu}_r - \bm{\mu}_g \|^2 + \operatorname{Tr}(\bm{\Sigma}_r + \bm{\Sigma}_g - 2 (\bm{\Sigma}_r \bm{\Sigma}_g)^{1/2})
\end{align}
where:
\begin{itemize}
    \item $\bm{\mu}_r$ and $\bm{\Sigma}_r$  represent the mean vector and covariance matrix of the real ECG feature distribution,
    \item $\bm{\mu}_g$ and $\bm{\Sigma}_g$ represent the mean vector and covariance matrix of the generated ECG feature distribution, and
    \item $\operatorname{Tr}$ denotes the trace of a matrix.
\end{itemize}

\subsection{Model Performance On Synthetic Signals}
\begin{figure}[htb!] 
\centering
\begin{subfigure}{0.9\textwidth}
  \centering
  \includegraphics[width=\linewidth]{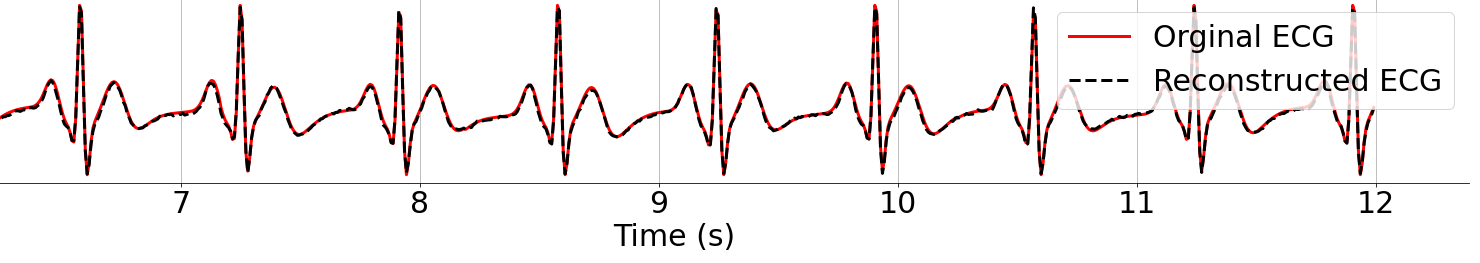}
  \caption{RSR rhythm.}
  \label{fig:synthetic_RSR}
\end{subfigure}
\begin{subfigure}{0.9\textwidth}
  \centering
    \includegraphics[width=\linewidth]{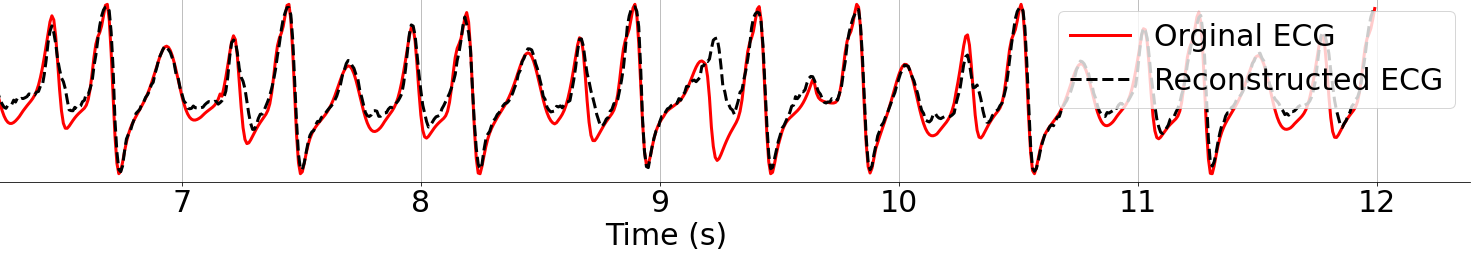}
  \caption{SA rhythm.}
  \label{fig:synthetic_SA} 
\end{subfigure}
\begin{subfigure}{0.9\textwidth}
  \centering
    \includegraphics[width=\linewidth]{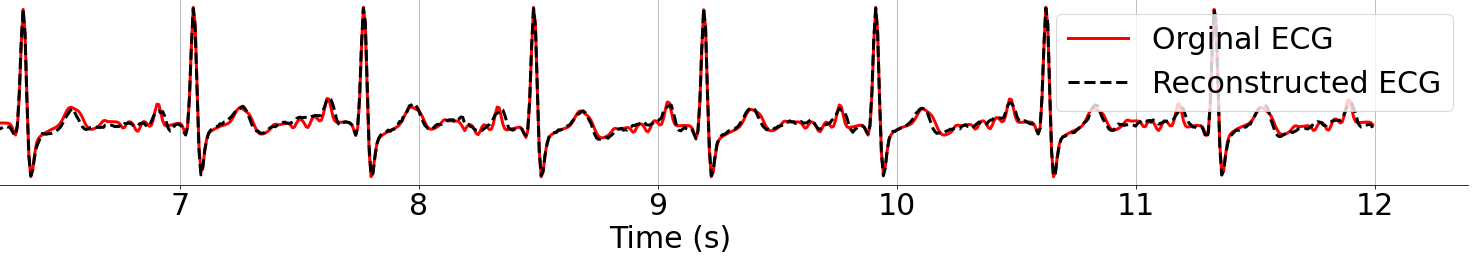}
  \caption{AFib rhythm.}
  \label{fig:synthetic_AF}
  \end{subfigure}
\caption{Reconstructed three rhythms obtained from our CLEP-GAN method.}
\label{fig:synthetic_results}.
\end{figure}

As detailed in Section \ref{sec:synthetic_data}, we employed our ODE model to generate three different ECG-PPG rhythms, each consisting of $34$ distinct ECG-PPG pairs. For the evaluation of our approach using the synthetic dataset, we randomly selected three signals from each rhythm as the testing data, while the remaining signals were assigned to the training set. Fig. \ref{fig:synthetic_results} displays reconstructed signals obtained through our CLEP-GAN method using the synthetic dataset.

In Fig. \ref{fig:synthetic_results}, the reconstructed ECG signals closely mirror the ground truth. This observation is supported by the quantitative results in Tables \ref{table:synthetic_RMSE_MAE_HRV_results} and \ref{table:synthetic_RR_dist_evaluation}. Notably, for both RSR and AFib rhythms, both the heart rate and heart rate variability are precisely retained. The RMSE values of the reconstructed ECG waveforms consistently remain below $0.1$, highlighting the effectiveness of the reconstruction process.

\begin{table}[htb!] 
\centering
\setlength{\tabcolsep}{4pt}
\begin{tabular}{@{}c|c|c|c|cc|cc@{}} 
 \toprule
  &  &  &  &\multicolumn{4}{c}{HRV}\\
 & & & & \multicolumn{2}{c|}{Reconstructed} & \multicolumn{2}{c}{Ground-truth} \\
 \multirow{-3}{*}{Rhythms} & \multirow{-3}{*}{Signal} & \multirow{-3}{*}{RMSE} & \multirow{-3}{*}{$MAE_{HR}$} & Mean & STD & Mean & STD \\

 \midrule
  & 7 & 0.02 & 0.0 & 665.45 & 3.09  & 665.45 & 3.09 \\ 
  & 16 & 0.03 & 0.0  & 545.71 & 3.28 & 545.71 & 3.28 \\ 
 \multirow{-3}{*}{RSR}  & 22 & 0.07 & 0.0 & 691.20 & 3.92 & 691.20 & 3.92 \\ 
 \midrule
  & 8 & 0.16 & 0.61  & 601.85 & 4.61  & 601.85 & 3.37\\ 
  & 30 & 0.11 & 0.40 & 664.73 & 8.67 & 664.73 & 9.92\\ 
 \multirow{-3}{*}{SA}  & 34 & 0.12 &  0.6 & 688.80 & 10.70 & 688.80 & 11.5\\ 
 \midrule
  & 9 & 0.06 &  0.0 & 781.60 & 3.67 & 781.60 & 3.67\\ 
  & 37 & 0.06 &  0.0 & 673.45 & 8.23 & 673.45 & 8.23 \\ 
 \multirow{-3}{*}{AFib.} & 42 & 0.04 &  0.0 & 712.80 & 4.31 & 712.80 & 4.31\\ 
 \bottomrule
\end{tabular}
\caption{Quantitative results of our method applied to three synthetic ECG-PPG rhythms. $MAE_{HR}$ is meansured in milliseconds.}
\label{table:synthetic_RMSE_MAE_HRV_results}
\end{table}

In contrast, the SA rhythm presents a marginally reduced accuracy across most evaluation metrics, particularly those assessing RR intervals. This deviation might be linked to the inherent waveform traits of the SA rhythm, especially the irregularities observed in the T waves. Such irregularities can elevate the signal's RMSE values and pose challenges to the peak finding algorithm, leading to potential misidentification of T waves as R peaks.

\begin{table}[htb!] 
\small
\centering
\begin{tabular}{cccccc} 
 \toprule
 Rhythms & rHI $\uparrow$  & rRMSE $\downarrow$ & rEMD $\downarrow$ & KS $\downarrow$ & KL $\downarrow$\\
 \midrule 
 RSR &  0.98 &  0.03 & 7.9e-3 & 8.22e-6 & 0.05\\ 
  \midrule
 SA & 0.94 & 0.02 & 0.02 & 5.04e-5 & 0.08 \\ 
 \midrule
 AFib. &  0.98 &  4.51e-3 &  6.01e-3 & 3.56e-5 & 0.06\\ 
 \bottomrule
\end{tabular}
\caption{Average discrepancies in RR intervals between CLEP-GAN reconstructed synthetic ECGs and their corresponding ground truth, as generated by our proposed ODE model. The NeuroKit algorithm is employed for peak detection.}
\label{table:synthetic_RR_dist_evaluation}
\end{table}

\subsection{Evaluation on Real Datasets} \label{sec:evaluation_on_real_dataset}
\begin{figure}[htb!] 
\centering
\begin{subfigure}{0.9\textwidth}
  \centering
  \includegraphics[width=\linewidth]{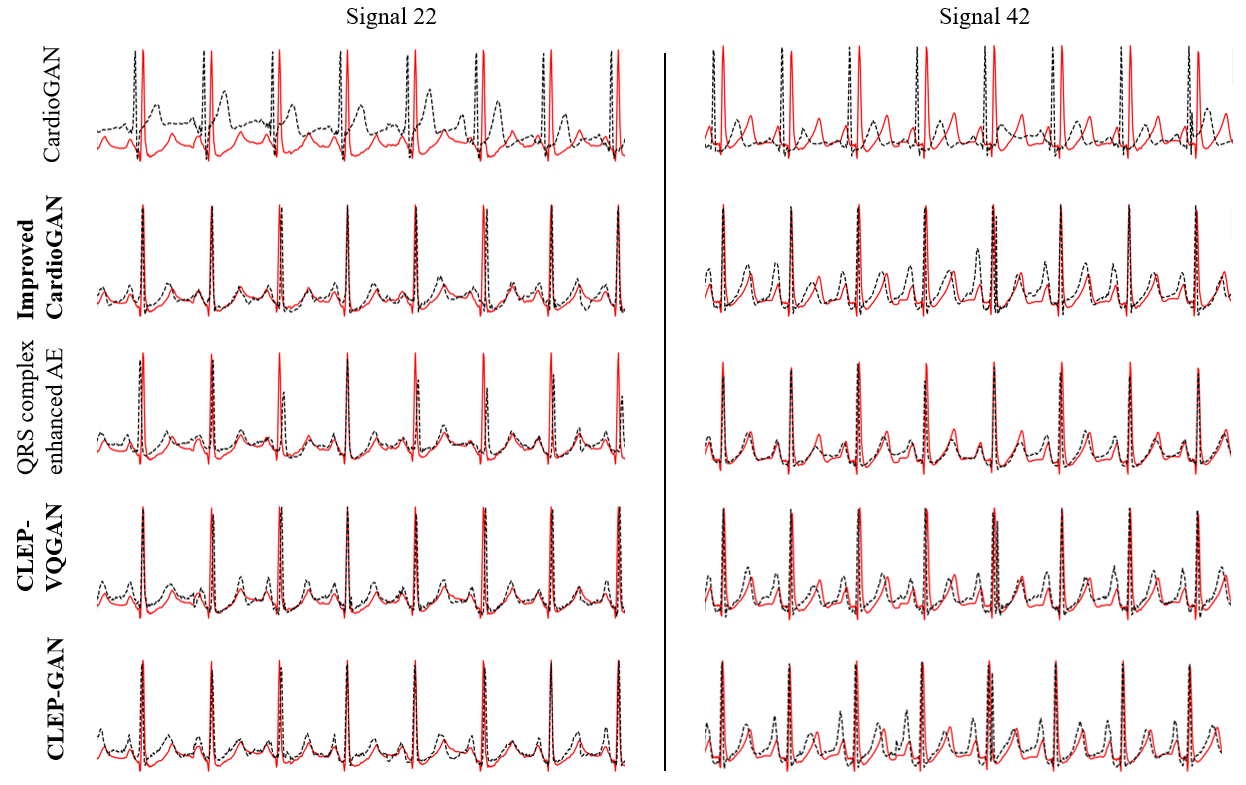}
  \caption{On BIDMC dataset.}
  \label{fig:comparison_visual_BIDMC}
\end{subfigure}
\begin{subfigure}{0.9\textwidth}
  \centering
    \includegraphics[width=\linewidth]{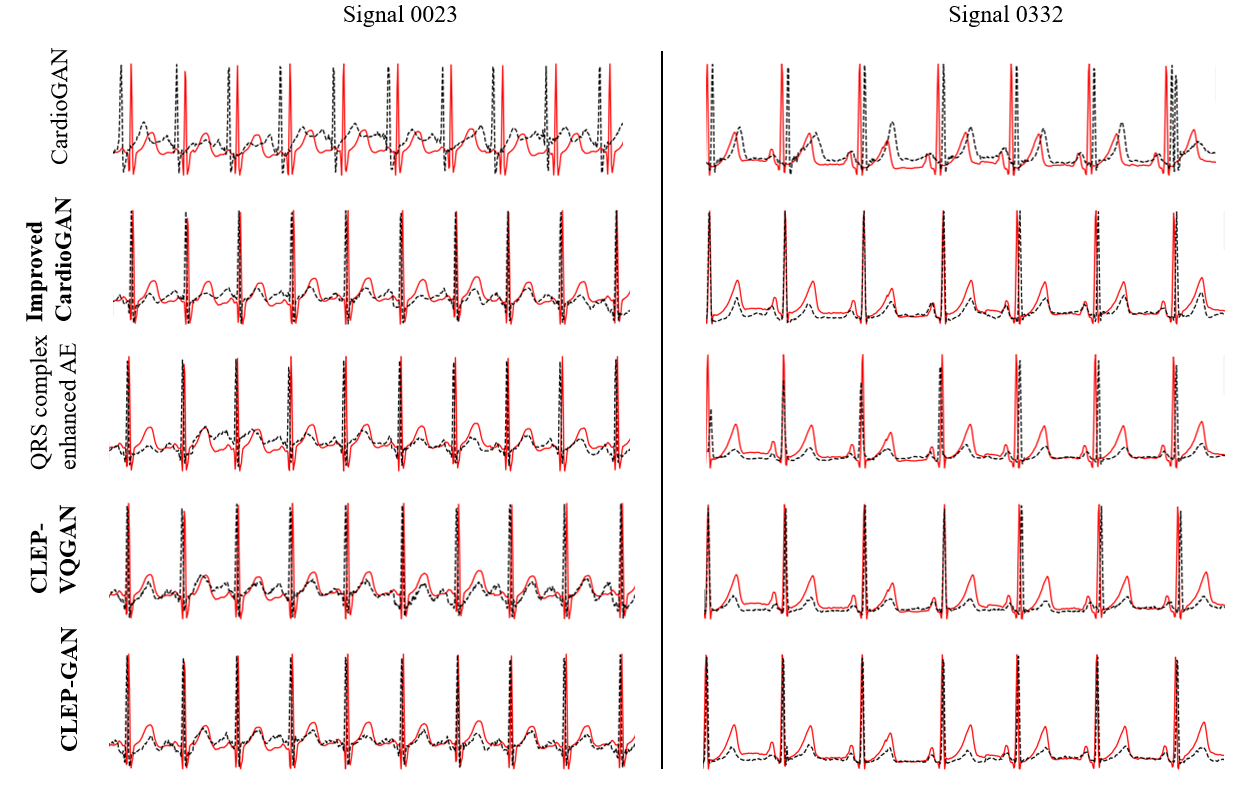}
  \caption{On CapnoBase dataset.}
  \label{fig:comparison_visual_CapnoBase}
\end{subfigure}
\caption{Reconstructed ECG samples in BIDMC and CapnoBase datasets of five methods: CardioGAN \cite{Sarkar2021}, our improved CardioGAN, QRS complex-enhanced encoder-decoder \cite{Chiu2020}, our CLEP-VQGAN and our proposed CLEP-GAN. In the graphical representation, the ground truth ECG traces are depicted with solid red lines, while the reconstructed ECG traces are illustrated with dashed black lines. Note: The implementations of the methods CardioGAN and QRS complex-enhanced encoder-decoder are based on the official code available in references \cite{cardiogan_git_2020} and \cite{QRS_enhanced_AE_2020}, respectively.}
\label{fig:reconstructed_real_ECGs}.
\end{figure}

As elaborated in Section \ref{sec:real_dataset}, our evaluation involves two real datasets: BIDMC and CapnoBase. In our current assessment, we have chosen $34$ ECG-PPG pairs with minimal noise from each dataset. To ensure a comprehensive evaluation, $15\%$ of the pairs from each dataset were randomly set aside for testing, with the rest earmarked for training. Fig. \ref{fig:reconstructed_real_ECGs} illustrates the reconstructed ECG signals from the testing set, offering a visual insight into the results. Tables \ref{table:results_RMSE_MAE}, and \ref{table:results_HRV} provide a quantitative comparison of our three introduced methods: improved CardioGAN (CardioGAN+), CLEP-VQGAN, and CLEP-GAN, against three established algorithms, namely CardioGAN \cite{Sarkar2021}, the QRS complex-enhanced encoder-decoder (QRS-ED.) \cite{Chiu2020}, and RDDM \cite{Shome2023}. Our implementation of the CardioGAN, QRS complex-enhanced encoder-decoder methods, and RDDM are based on the official code made available in references \cite{cardiogan_git_2020},  \cite{QRS_enhanced_AE_2020}, and \cite{Shome2023git}, respectively.

In our improved CardioGAN implementation, we retain the core architecture of the original CardioGAN, with several key modifications. We replace the $L_1$ loss function with smooth $L_1$ for the cyclic consistency (or reconstruction) loss and introduce the mid-way reconstruction loss, represented by the first term in the following $L_{recon}$ function. Additionally, we have simplified the process to a single loop, $P \rightarrow G_E(P) \rightarrow G_P(G_E(P)) \rightarrow P'$, instead of the original two cycles. The reconstructed loss is given by the equation:
\begin{align}
    L_{recon}(G_E, G_P) &= \mathbb{E}_{p \sim P}[||G_E(p) - e||_{\text{smooth\_{$L_1$}}}] \nonumber \\
    & + \mathbb{E}_{e' \sim E'}[||G_P(e') - p||_{\text{smooth\_{$L_1$}}}],
\end{align}
Where $E'$ and $P'$ are reconstructed ECG and PPG, respectively.

As shown in Table  \ref{table:results_RMSE_MAE}, evaluating the performance with RMSE reveals that the original CardioGAN falls short compared to other methods. Additionally, Fig. \ref{fig:reconstructed_real_ECGs} demonstrates that the ECGs reconstructed by CardioGAN deviate from the ground truth, with a noticeable shift in the QRS complex. This shift largely contributes to its higher RMSE, despite maintaining a reasonable mean absolute error in heart rate estimation. Comparing CardioGAN with our enhanced version (CardioGAN+), we observe an approximate 10\% reduction in average RMSE (see Table \ref{table:results_RMSE_MAE}), with minimal impact on the accuracy metrics for $MAE_{HR}$ and HRV across both datasets. When assessing RR interval distributions, each technique displays unique strengths depending on the dataset used.

Our CLEP-GAN method manifests impressive outcomes, particularly on the CapnoBase dataset in both RMSE and FD. Diving deeper, the method from \cite{Chiu2020} has a slightly lower RMSE on the BIDMC dataset compared to our CLEP-GAN. However, its mean $MAE_{HR}$ is notably higher, standing at $1.75$, nearly double the $0.84$ achieved by the CLEP-GAN. A visual inspection as depicted in Fig. \ref{fig:reconstructed_real_ECGs}, indicates that the ECG waveforms reconstructed by CLEP-GAN more accurately mirror the ground truths, a fact especially evident for testing signal 22 from the BIDMC dataset and testing signal 0332 from the CapnoBase dataset.

In our experiments, diffusion-based RDDM doesn't show any advantages. One potential reason might be that we set the training epoch to 500 rather than the default 1000 epochs used in their published code, considering computational efficiency. Additionally, a limitation of RDDM is the testing time; even though the authors of RDDM limited it to 10 diffusion steps, the testing time is still nearly three times longer than that of other models.

\begin{table}[htb!] 
\centering
\begin{tabular}{@{}c|c|c|c|c@{}} 
 \toprule
Dataset & Method & RMSE $\downarrow$ & FD $\downarrow$ & $MAE_{HR}$ $\downarrow$ \\
\midrule                
                        & CardioGAN & 0.47 & 34.29 &  \textbf{0.60}\\
                        & CardioGAN+ (ours) & 0.37 & 25.38 & 2.28 \\
                        & QRS-ED. &  \textbf{0.36} & 24.71 & 1.75 \\
                        & RDDM &      0.38    & 27.46 & 1.07\\
                        & CLEP-VQGAN (ours) & 0.37 & \textbf{22.10} & 0.89 \\
\multirow{-4}{*}{BIDMC} & CLEP-GAN (ours)& 0.37 & 22.27 & 0.84\\
\midrule  
                        & CardioGAN & 0.45 & 47.83 & 1.27\\
                        & CardioGAN+ (ours) & 0.34 & 33.87 & \textbf{1.02} \\
                        & QRS-ED. & 0.36 & 32.49 & 1.71 \\
                        & RDDM &    0.36  & 36.08 & 1.27\\
                        & CLEP-VQGAN (ours) & 0.35 & 35.01 & 1.54 \\
\multirow{-4}{*}{CapnoBase} & CLEP-GAN (ours)& \textbf{0.33} & \textbf{32.45} & 1.29\\

\bottomrule
\end{tabular}
\caption{Comparison of RMSE, FD and $MAE_{HR}$ between our methods (i.e., improved CardioGAN (CardioGAN+), CLEP-VQGAN, and CLEP-GAN) and other advanced methods: CardioGAN, QRS complex-enhanced encoder-decoder (QRS-ED.), and RDDM,  across two datasets: BIDMC and CapnoBase. The training epochs are set to 200 for improved CardioGAN (CardioGAN+), QRS-ED, and CLEP-GAN, and 500 for CLEP-VQGAN and RDDM.}
\label{table:results_RMSE_MAE}
\end{table}

Experimental results demonstrate that the Attention U-Net generator outperformed the VQ-VAE model in evaluations on two real datasets. One reason may be that VQ-VAE quantizes input data into discrete latent representations using a fixed codebook. If the codebook size is insufficient, it may lead to information loss, as the limited number of entries cannot fully capture the variability and fine details of the input data. This limitation is problematic for tasks like ECG reconstruction, where accurately representing subtle waveform details is crucial. Increasing the codebook size could reduce this information loss by providing more options for capturing details. However, larger codebooks also increase memory requirements and computational complexity, which may lead to overfitting on the training data if the size becomes excessive.

\begin{figure}[htb!] 
\centering
\begin{subfigure}{0.9\textwidth}
  \centering
    \includegraphics[width=\linewidth]{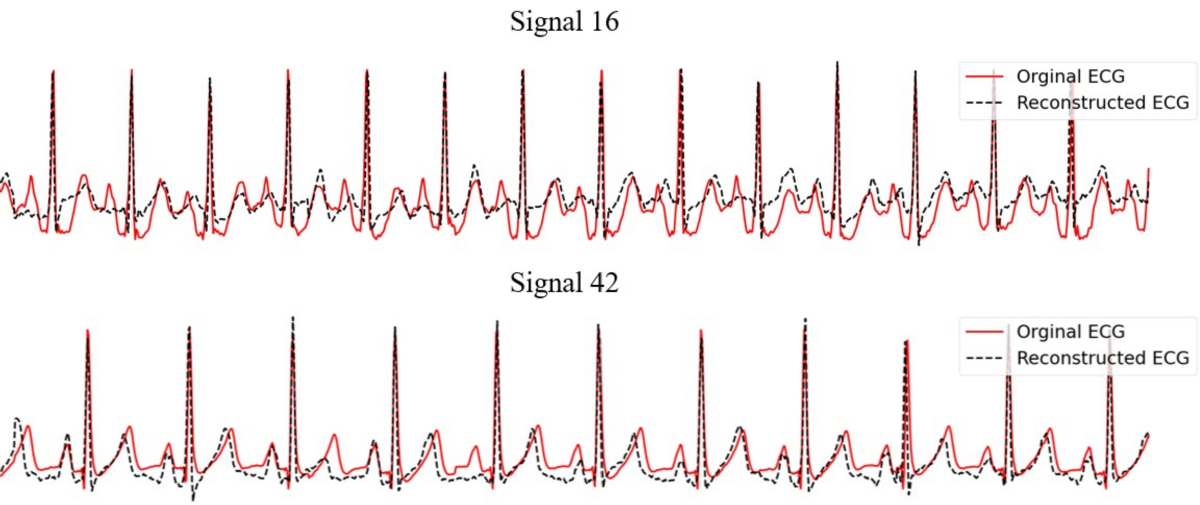}
  \caption{VQ-VAE-based CLEP-VQGAN.}
  \label{fig:VQ-VAE2} 
\end{subfigure}
\begin{subfigure}{0.9\textwidth}
  \centering
    \includegraphics[width=\linewidth]{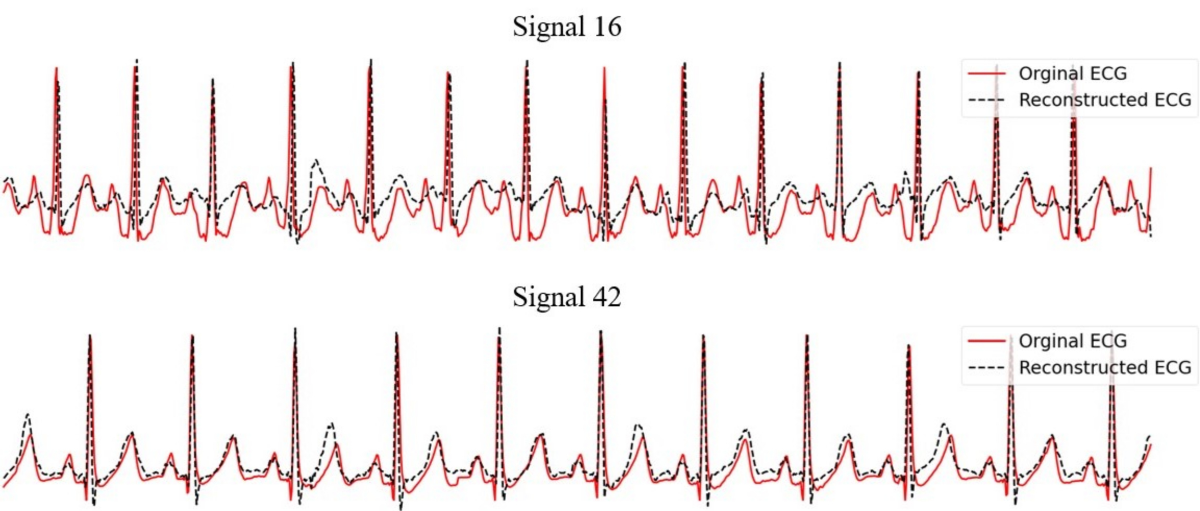}
  \caption{ VQ-VAE2-based CLEP-VQGAN.}
  \label{fig:VQ-VAE2}
  \end{subfigure}
\caption{A visual comparison between VQ-VAE-based CLEP-VQGAN and VQ-VAE2-based CLEP-VQGAN across two signals.}
\label{fig:VQVAEvsVQVAE2}.
\end{figure}

In all experiments, we use samples with a sequence length of 512 points. Given a sampling rate of 125 Hz, each input sample spans approximately 4 seconds. To investigate whether VQ-VAE’s lower performance was related to sequence length limitations, we replaced VQ-VAE with VQ-VAE2 \cite{razavi2019vqvae2}, a hierarchical model designed to handle longer sequences more effectively and to capture different levels of detail. However, experimental results show that while VQ-VAE2 performed better than the original VQ-VAE on some signals, it did not demonstrate consistent improvement across all testing signals. One possible reason may be that, although VQ-VAE2 captures more detail, it is also more sensitive to signal noise. Fig. \ref{fig:VQVAEvsVQVAE2} presents a visual comparison between VQ-VAE-based CLEP-VQGAN and VQ-VAE2-based CLEP-VQGAN across two signals. For the noisy signal (signal 16), the VQ-VAE-based CLEP-VQGAN achieves a lower RMSE of $0.45$ compared to $0.47$ for the VQ-VAE2-based CLEP-VQGAN. In contrast, for the less noisy signal (signal 42), the VQ-VAE2-based CLEP-VQGAN yields a better RMSE of $0.27$, compared to $0.29$ for the VQ-VAE-based CLEP-VQGAN.

\begin{table}[htb!] 
\centering
\small
\setlength{\tabcolsep}{3pt}
\begin{tabular}{@{}c|ccc|ccc@{}} 
 \toprule
 & \multicolumn{3}{c|}{BIDMC} & \multicolumn{3}{c}{CapnoBase} \\
  \multirow{-2}{*}{Method} & Signal & Mean & STD & Signal & Mean & STD\\
 \midrule
&  07 &  664.67/665.45 &  3.94/5.73 &   0018 &  424.44/424.00 &  3.24/6.53 \\ 
&  16 &  545.85/546.29 &  3.37/5.60 &  0023  &  \textbf{560.00/560.00} &  8.88/16.00 \\ 
&  22 &  718.40/717.82 &  4.80/7.70 &   0104 &  \textbf{512.00/512.00} &  0.01/6.53 \\  
\multirow{-4}{*}{CardioGAN} & 42 & 712.00/712.80 &  5.06/4.31 &  0332 &   \textbf{807.11/807.11} &  23.69/26.52\\       
\midrule
&  07 &  664.67/665.45 &  3.94/7.49 &  0018 &  \textbf{424.44/424.44} &  3.24/6.78 \\ 
&  16 &  545.85/549.67 &  3.37/50.09 &  0023 &  \textbf{560.00/560.00} &   8.88/11.74 \\ 
\multirow{-2}{*}{CardioGAN+} &  22 &  718.40/718.40 &   4.80/20.18 &  0104 &   \textbf{512.00/512.00} &  0.01/4.28 \\
\multirow{-2}{*}{(ours)} &  42 &  712.00/711.20 &  5.06/5.60 & 0332 &  807.11/808.89 & 23.69/0.84 \\         
\midrule
&  07 &  664.67/664.00 &  3.94/14.06 &  0018 &  424.44/425.33 &  3.24/12.00 \\ 
&  16 &  545.85/545.23 &  3.37/9.33 &  0023 &  560.00/558.77 &   8.88/11.25\\ 
&  22 &  \textbf{718.40/718.00} &   4.80/32.36 &  0104 &  512.00/511.43 &  0.01/5.63\\ 
\multirow{-2}{*}{QRS-ED.} &  42 & \textbf{712.00/712.00} & 5.06/10.12 & 0332 &  807.11/809.78 &  23.69/28.42 \\ 
\midrule
&  07 &  \textbf{664.67/664.73}  &  3.94/7.97 &  0018 &  424.44/425.78 &  3.24/10.85\\ 
&  16 &  \textbf{545.85/545.85} &  3.37/6.40 &  0023 &   560.00/559.38 &   8.88/12.34\\ 
\multirow{-2}{*}{CLEP+VQGAN} &  22 &  718.40/716.00 &   4.80/20.94 &  0104 &   \textbf{512.00/512.00} &  0.01/9.07 \\  
\multirow{-2}{*}{(ours)} & 42 & \textbf{712.00/712.00} & 5.06/5.06 & 0332 &  807.11/808.89 & 23.69/20.81 \\ 
\midrule
&  07 &  664.67/665.45 &  3.94/5.73 &  0018 &  424.44/424.89 &  3.24/7.00 \\ 
&  16 &  \textbf{545.85/545.85} &  3.37/5.57 &  0023 &   560.00/559.38 &   8.88/12.34 \\ 
\multirow{-2}{*}{CLEP-GAN} &  22 &  718.40/716.80 &   4.80/15.68 &  0104 &   \textbf{512.00/512.00} &  0.01/7.41 \\ 
\multirow{-2}{*}{(ours)} & 42 & \textbf{712.00/712.00} &  5.06/6.20 & 0332 &   807.11/808.00 & 23.69/20.67\\ 
\bottomrule
\end{tabular}
\caption{Comparison of HRV (i.e., mean of RR interval and standard derivation of RR interval) between our methods and two other advanced methods. RR interval is measured in milliseconds. The results are presented as ground truth/reconstruction.}
\label{table:results_HRV}
\end{table}

\subsubsection{T Wave Reconstruction}
Accurately reconstructing T waves remains a challenging and an open question. Studies such as \cite{Elgendi2012, Charlton2016, Zhu2021} discuss the difficulties in precisely mapping specific waves, including the T wave, due to inherent limitations in PPG signals. On the other hand, studies like \cite{Nishimura1986, Kachuee2015, Pimentel2017, Charlton2017} suggest that certain aspects of PPG may indirectly reflect components of the cardiac cycle typically associated with the ECG T wave. Notably, \cite{Zhu2021} is among the few studies that propose the potential for partial reconstruction of T wave characteristics from PPG, though acknowledging that the accuracy may be limited.

\begin{figure}[htb!] 
\centering
\includegraphics[width=\linewidth]{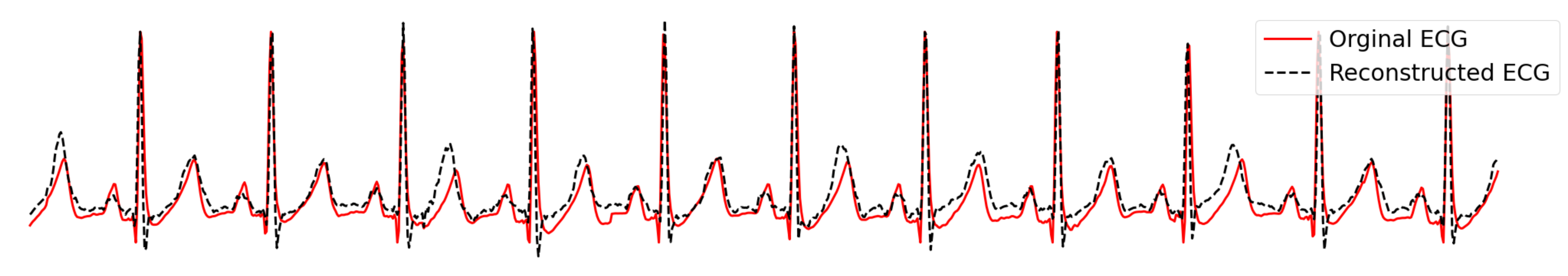}
\caption{Sample reconstructed by the CLEP-VQGAN method.}
\label{fig:CLEP_VQGAN_sample42}
\end{figure}

In our study, we further explore this question by aiming to reconstruct the full ECG waveform, rather than focusing solely on heart rate variability (HRV) or heart rate. Our experiments suggest that partial reconstruction of the T wave is possible, as shown in Fig. \ref{fig:CLEP_VQGAN_sample42}, where the T wave is successfully reconstructed in certain heartbeats. However, consistently locating and accurately reconstructing T waves remains challenging.

To assess reconstruction accuracy across all ECG components, including the T wave, we use common point-to-point metrics such as RMSE. Although our method shows improved overall performance compared to other approaches, accurately reconstructing smaller waves like the T wave remains a challenge. In future work, we plan to focus on specific features of these small waves, such as amplitude and width, incorporating them into the input data to enhance model learning. Addressing these details will be a key priority as we continue to explore the complexities of ECG waveform reconstruction from PPG.

\subsection{Complexity}
The computational complexity of CNNs is commonly measured by the number of Floating Point Operations (FLOPs) and the total parameter count. FLOPs estimate the computational effort required for forward inference, while parameters quantify memory requirements. For a convolutional layer, the FLOPs are computed as $ H_{\text{out}} \times W_{\text{out}} \times C_{\text{out}} \times C_{\text{in}} \times K_h \times K_w $, where $H_{\text{out}} $ and $ W_{\text{out}} $ are the output dimensions, $ C_{\text{in}} $ and $ C_{\text{out}} $ are the input and output channels, and $ K_h $ and $ K_w $ denote the kernel dimensions. In the context of a 1D convolution, Height $H$ and $K_w$ are set as $1$. Parameters for convolutional layers are given by $ C_{\text{out}} \times C_{\text{in}} \times K_h \times K_w + C_{\text{out}} $, including biases. Fully connected layers add complexity linearly with $\text{FLOPs} = N_{\text{in}} \times N_{\text{out}}$ and $ \text{Parameters} = N_{\text{in}} \times N_{\text{out}} + N_{\text{out}} $, where $ N_{\text{in}} $ and $ N_{\text{out}} $ are the number of input and output features, respectively.

Pooling and normalization layers introduce minimal computation. Pooling FLOPs depend only on the output dimensions, while normalization adds $ 2 \times H_{\text{out}} \times W_{\text{out}} \times C $ FLOPs for scaling and shifting, with parameters totalling $ 2 \times C $. For attention layers, common in transformers, the FLOPs increase with feature and head dimensions: $ \text{FLOPs} = 3 \times d^2 \times n_h + d^2 $ and $\text{Parameters} = 4 \times d^2 $, where $ d $ is the token dimension, and $ n_h $ is the number of heads. 

Table \ref{table:complexity} presents the computational complexity, training time, and testing time for each method. Compared to CardioGAN, our approach demonstrates both lower time and space complexity. The training time of our method is shorter than that of both CardioGAN and the diffusion-based RDDM model. During testing, CardioGAN and our enhanced version (CardioGAN+) have significantly lower testing times than other models, as they utilize a single Attention U-Net to generate an ECG signal directly from a PPG input. In contrast, our CLEP-GAN model requires using the encoder part of the PPG generator (Attention U-Net) to produce a latent representation, which is then input into the decoder part of the ECG generator to reconstruct the ECG signal. This extra step adds an increase in processing time.

All experiments were conducted on a machine equipped with an NVIDIA Quadro RTX 6000 GPU with 24 GB of memory.

\begin{table}[htb!] 
\centering
\small
\setlength{\tabcolsep}{3pt}
\begin{tabular}{@{}cc|cc|c|c|c@{}} 
\toprule
&  & \multicolumn{2}{c|}{Complexity} &  & \\
\multicolumn{2}{c|}{Method} & Time Comp. & Space Comp. & Train Time & Testing Time & Performance \\
&                           & (MFLOPs) & (MParams) & (Sec.) & (Sec.) & (FD)\\
\midrule
CardioGAN                   & A-UNet  &           &        &        &  \\ 
                            &  (w/o CLIP) & $2 \times 1371.46$ & $2 \times 36.19$  & &  \\ 
                            &  TD  & $2 \times 117.72$  & $2 \times 2.76$ &  &  \\ 
                            &  FD  & $2 \times 14.08$  & $2 \times 1.21$ &  &  \\ 
                            &  Total & 3006.52 & 80.32 & 248.71 & \textbf{0.23} & 34.29\\ 
\midrule   
CardioGAN+                  & A-UNet   &       &       &         &  \\ 
(ours)                      & (w/o CLIP) & $2 \times 1371.46$ & $2 \times 36.19$  & &  \\ 
                            &  TD  & 117.72  &  2.76 &  &  \\ 
                            &  FD  & 14.08  &  1.21 &  &  \\ 
                            &  Total & 2874.72 & 76.35& 122.45 & \textbf{0.23} & 25.38 \\ 
\midrule   
&  &  &  &  &  \\ 
QRS-ED.                    &  & 1165.60 & \textbf{11.34} & \textbf{78.47}  & 1.08 & 24.71\\ 
\midrule   
RDDM                       &   RDDM net  & 544.87  & 22.77 &  &  \\ 
                           &  Cond. net & $2 \times 874.09$ & $2 \times 26.94$ & &  \\ 
                           &  Total  & 2293.05 & 76.65 & 193.84 & 3.84 & 27.46\\ 
\midrule
CLEP-VQGAN                 & VQVAE & $2 \times 39.85$ & $2 \times 11.99$ &   & \\ 
(ours)                     & TD     & 117.72  & 2.76 &  &  \\ 
                           & FD  & 14.08  & 1.21 &  &  \\ 
                           & Total  &  \textbf{211.50} & 27.95 & 117.34 & 1. 07 & \textbf{22.10} \\ 
\midrule
CLEP-GAN                   &  A-UNet &  &  &  &  \\ 
(ours)                     &  (w/ CLIP) & $2 \times 1373.3$ & $2 \times 38.03$ & &  \\ 
                           &  TD  & 117.72  & 2.76 &  &  \\ 
                           &  FD  & 14.08  & 1.21 &  &  \\ 
                           &  Total   & 2878.4 & 80.03 & 166.39 & 1.03 & 22.27\\ 
\bottomrule
\end{tabular}
\caption{The models were tested on a single ECG-PPG pair over one epoch with a batch size of 32 and a sample length of 512. Incorporating two identical networks in the framework increases complexity to $2 \times \#$, doubling that of a single network. Unlike CardioGAN, which uses two dual discriminators for both PPG and ECG reconstructions, we employ only one dual discriminator for ECG generation. Note that, during the inference stage, both our proposed method and CardioGAN use only one encoder and one decoder of the Attention U-Net. However, during training, two Attention U-Nets and dual discriminators are utilized. MFLOPs: Million Floating Point Operations. MParams: Million Parameters. Sec: Seconds. FD: Fr\'{e}chet Distance.}
\label{table:complexity}
\end{table}

\subsection{Ablation Experiments}
Our proposed CLEP-GAN comprises two primary components: contrastive learning and adversarial learning, as elaborated in section \ref{sec:method_architecture}. To assess the significance of these two techniques, we conduct an ablation study on two variants of the CLEP-GAN method in this section.

Initially, we evaluate without the inclusion of contrastive learning to gauge its impact. In this configuration, the generator network directly takes the PPG as input and maps it to the ECG, but without employing contrastive learning. However, the dual discriminators for adversarial learning are still maintained. Secondly, we exclude adversarial learning by removing both the time and frequency domain-based discriminators.

Table \ref{table:ablation} presents the quantitative outcomes of the ablation study. It can be observed that eliminating either the contrastive learning or the adversarial learning results in an RMSE increase of about $2\%$ on the CapnoBase dataset. Interestingly, there is a slight decrease in $MAE_{HR}$ when contrastive learning is not employed. For the BIDMC dataset, the RMSE values across the three methods remain consistent. However, when the model operates without contrastive learning, the $MAE_{HR}$ is significantly higher compared to the other two methods.

Integrating contrastive learning into the Attention U-Net-based network increases the time complexity (FLOPs) only slightly, from 1371.46 million to 1373.3 million, and the space complexity from 36.19 million to 38.03 million, which is not a significant increase. Adding dual discriminators increases both time and space complexities by 121.7 million and 3.97 million, respectively. However, given the total complexities (2878.4 million and 80.03 million), this increase is acceptable. Thus, whether to include adversarial learning depends on the specific requirements of the real-world application, whether it prioritizes accuracy or computational efficiency.

\begin{table}[htb!]
\centering
\begin{tabular}{@{}c|cc|cc@{}} 
 \toprule
  & \multicolumn{2}{c|}{BIDMC} & \multicolumn{2}{c}{CapnoBase} \\
  \multirow{-2}{*}{Method}  & RMSE & $MAE_{HR}$ & RMSE & $MAE_{HR}$\\
 \midrule
  CLEP-GAN w/o &  & &  &  \\ 
  contrastive learning & \multirow{-2}{*}{0.37} & \multirow{-2}{*}{2.14} & \multirow{-2}{*}{0.35} & \multirow{-2}{*}{\textbf{1.02}} \\
 \midrule
  CLEP-GAN w/o &  &  &  &  \\ 
  adversarial learning & \multirow{-2}{*}{0.37} & \multirow{-2}{*}{\textbf{0.70}} & \multirow{-2}{*}{0.35} & \multirow{-2}{*}{1.29} \\ 
 \midrule
  CLEP-GAN  & & & & \\ 
  (proposed) & \multirow{-2}{*}{0.37} & \multirow{-2}{*}{0.84} & \multirow{-2}{*}{\textbf{0.33}} & \multirow{-2}{*}{1.29}\\ 
 \bottomrule
\end{tabular}
\caption{Quantitative results of ablation study.}
\label{table:ablation}
\end{table}

\subsection{Attention Map Visualization}
\begin{figure}[htb!] 
\centering
\begin{subfigure}{0.3\textwidth}
  \centering
   \includegraphics[width=\linewidth]{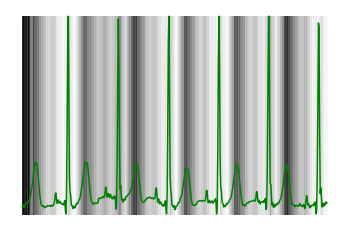}
  \caption{}
  \label{fig:atten_map_51}
\end{subfigure}
\begin{subfigure}{0.3\textwidth}
  \centering
    \includegraphics[width=\linewidth]{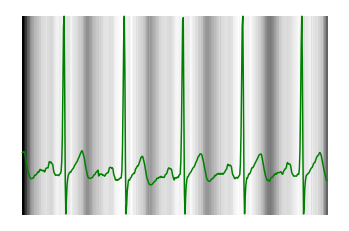}
  \caption{}
  \label{fig:atten_map_50} 
\end{subfigure}
\begin{subfigure}{0.3\textwidth}
  \centering
    \includegraphics[width=\linewidth]{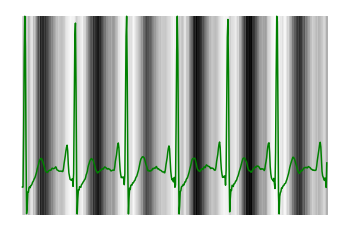}
  \caption{}
  \label{fig:atten_map_43}
  \end{subfigure}
\begin{subfigure}{0.3\textwidth}
  \centering
   \includegraphics[width=\linewidth]{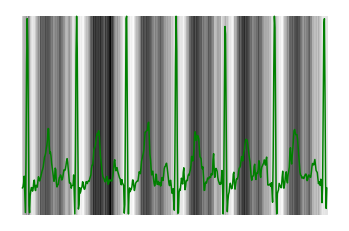}
  \caption{}
  \label{fig:atten_map_37}
\end{subfigure}
\begin{subfigure}{0.3\textwidth}
  \centering
    \includegraphics[width=\linewidth]{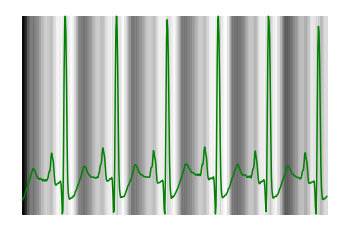}
  \caption{}
  \label{fig:atten_map_30} 
\end{subfigure}
\begin{subfigure}{0.3\textwidth}
  \centering
    \includegraphics[width=\linewidth]{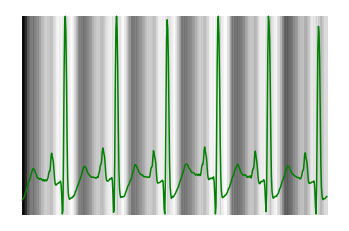}
  \caption{}
  \label{fig:atten_map_22}
  \end{subfigure}
\caption{Attention maps corresponding to six different signals from the BIDMC dataset. Brighter areas indicate regions where the generator pays more attention compared to the darker regions.}
\label{fig:attention_maps}.
\end{figure}

To better understand the role of the attention gate in emphasizing features crucial for ECG reconstruction, we visualized the attention maps applied to the last skip connection of the Attention U-Net generator ($G_E$). Fig. \ref{fig:attention_maps} overlays these attention maps onto the corresponding ground truth ECGs of six signals using our CLEP-GAN method. As depicted in this figure, the model prioritizes the QRS complex but gives less focus to the T wave. This pattern is consistent with our reconstruction results: whereas QRS reconstructions are impressive, challenges arise when reconstructing waves, particularly the T waves. A possible explanation for this might be the inherent nature of the T-wave. Unlike the sharply defined R-peak, the T-wave can exhibit significant variability in amplitude, shape, and duration, even within an individual. Factors such as heart rate, electrolyte balances, and medications can influence this variability, making the T-wave more challenging to correlate with features in the PPG.

\subsection{Exploring Influential Factors in ECG Reconstruction} \label{sec:exploring_influential_factors}
\subsubsection{Experimentation on a Single Subject} \label{sec:experimentation_on_a_Single_Subject}
\begin{figure}[htb!] 
\centering
\begin{subfigure}{0.9\textwidth}
  \centering
  \includegraphics[width=\linewidth]{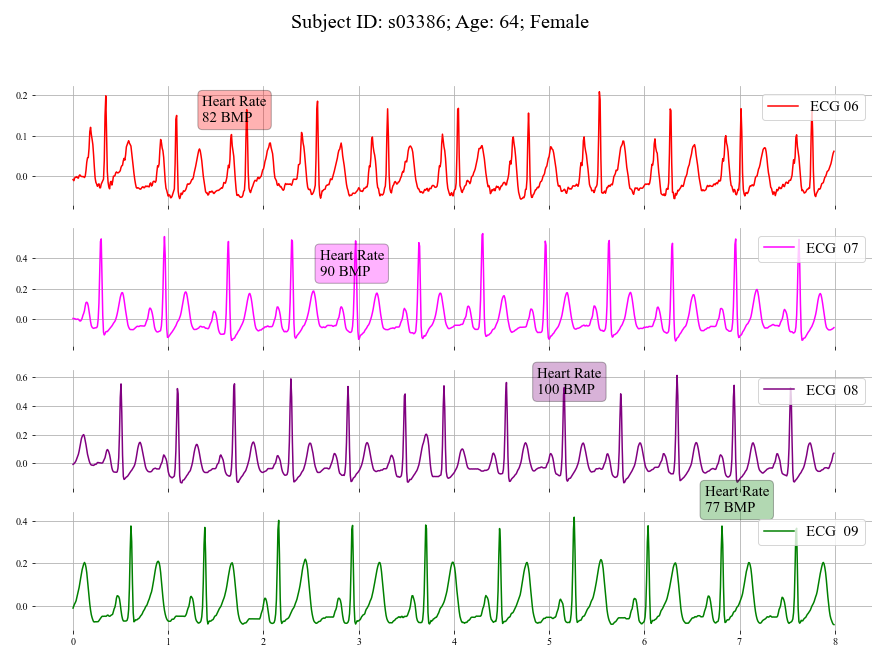}
  \caption{}
  \label{fig:s03386}
\end{subfigure}
\begin{subfigure}{0.9\textwidth}
  \centering
 \includegraphics[width=\linewidth]{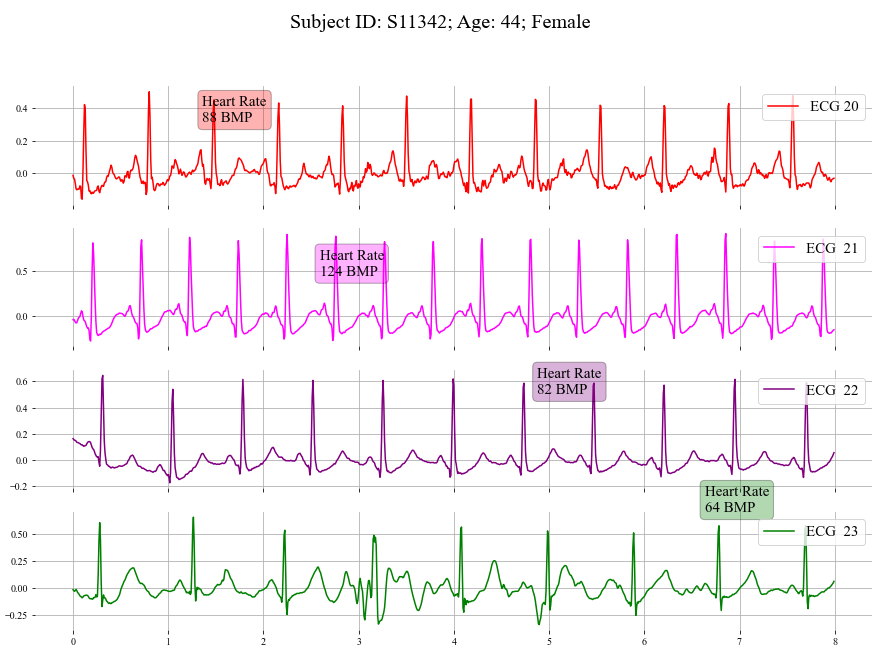}
  \caption{}
  \label{fig:s11342} 
\end{subfigure}
\caption{Illustration of the same patient has multiple waveforms in the BIDMC dataset.}
\label{fig:same_subject_multiple_ecgs}.
\end{figure}

The BIDMC dataset, a subset of the MIMIC II matched waveform database, contains prolonged patient monitoring, often resulting in multiple waveform records for an individual. These records represent various clinical episodes or time points. Fig. \ref{fig:same_subject_multiple_ecgs} shows how one patient can have diverse ECG rhythms over time. For instance, subjects s03386 and s11342 have four unique ECG-PPG pairs, revealing cardiac rhythm variability.

\begin{figure}[htb!] 
\centering
\begin{subfigure}{0.9\textwidth}
  \centering
   \includegraphics[width=\linewidth]{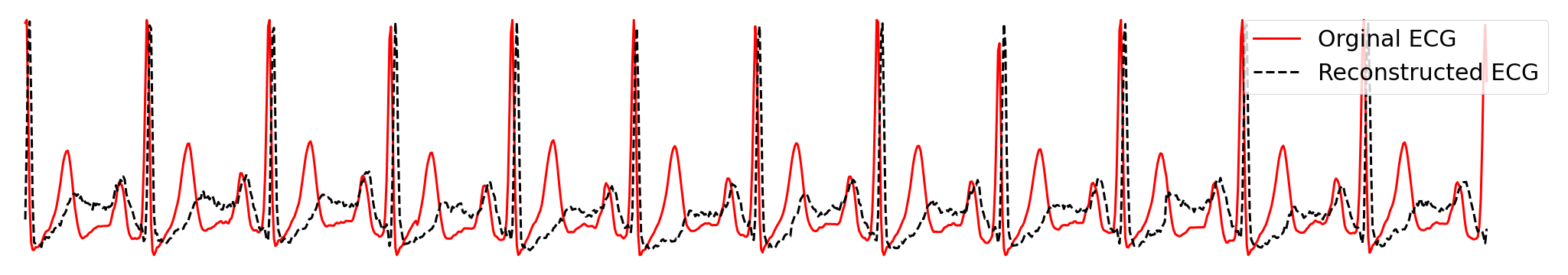}
  \caption{}
  \label{fig:s03386_07_all}
\end{subfigure}
\begin{subfigure}{0.9\textwidth}
  \centering
    \includegraphics[width=\linewidth]{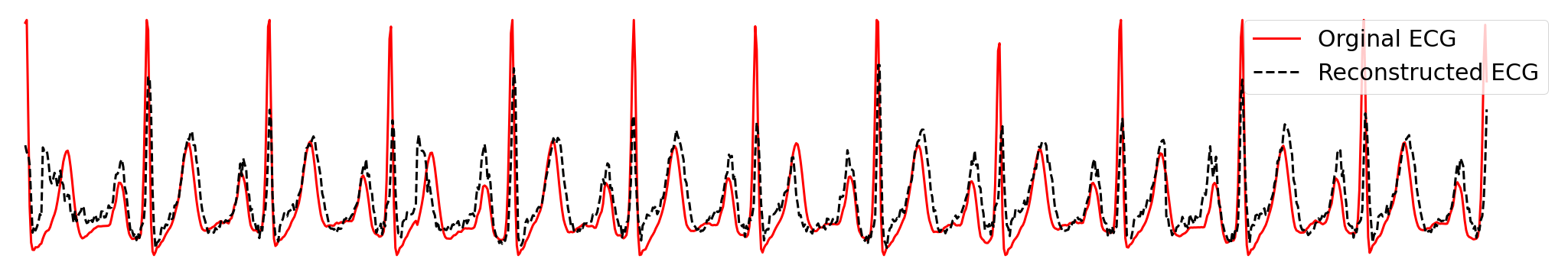}
  \caption{}
  \label{fig:s03386_07_three} 
\end{subfigure}
\begin{subfigure}{0.9\textwidth}
  \centering
    \includegraphics[width=\linewidth]{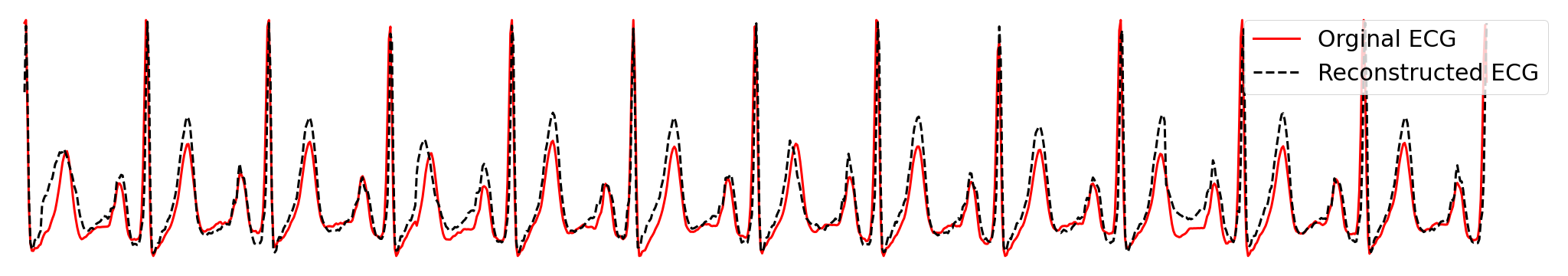}
  \caption{}
  \label{fig:s03386_07_pretrain}
  \end{subfigure}
\caption{
Comparison of results achieved with the proposed CLEP-GAN method under three training set conditions, with the same testing signal 07. (a) depicts the ECG reconstruction resulting from training the model on a more extensive dataset, comprising 42 ECG-PPG pairs from 35 subjects available in the BIDMC dataset. (b) displays the ECG reconstruction attained through training the model using three specific ECG-PPG pairs: signal 06, signal 08, and signal 09. These signals, along with the testing signal 07, originate from the same subject (ID: s03386). (c) illustrates the outcome of a two-step training process: initial pretraining on our synthetic dataset followed by fine-tuning using signals 06, 08, and 09.}
\label{fig:s03386_results}.
\end{figure}

While the methodologies previously discussed excel in R peak reconstruction, they struggle with other waveforms, as shown in Fig. \ref{fig:s03386_07_all}. This is due to factors like noise, sex \cite{surawicz2003differences, ezaki2010gender, ravens2018sex, Carbone2020, Zhu2021}, age \cite{Zhu2021}, and varying health conditions. Thus, creating a semantic space that accurately represents all ECGs is challenging.

In our experiment, rather than training on a broad dataset, we focused on individual patient signals, potentially reducing age and sex influences on ECG reconstruction. For subject s03386, we selected one ECG-PPG pair from the available four pairs as the testing signal, reserving the remaining three pairs for training. To improve accuracy, we employed transfer learning techniques, first training the model on our synthetic dataset and then fine-tuning it using the selected ECG-PPG training pairs.

Fig. \ref{fig:s03386_results} shows the reconstructed ECGs when CLEP-GAN is trained with different ECG-PPG pairs. A closer look at Fig. \ref{fig:s03386_07_all} and Fig. \ref{fig:s03386_07_three} reveals that, while the model trained on the full BIDMC dataset reconstructs the R peaks reasonably well, it exhibits some misalignment and does not accurately capture the T waves. In contrast, training the model on just three signals from the same patients improves the accuracy of the ECG waveform reconstruction, though there is a noticeable reduction in the amplitude of the R peaks compared to the ground truth ECG. Furthermore, as shown in Fig. \ref{fig:s03386_07_pretrain}, pretraining the model on our synthetic dataset followed by fine-tuning on separate signals from the same individual produces waveforms that more closely resemble the ground truth, with slightly elevated T wave amplitudes.

To prepare for pretraining the model, which is then fine-tuned on real data, we generated a large set of synthetic data using our ODE method, based on ECG-PPG pairs from the BIDMC dataset. We first selected 18 low-noise pairs and crafted three parameters: $\bm{a}$, $\bm{b}$, and $\bm{\theta}$, representing amplitude, width, and reference angles, respectively. For each pair, we varied the heartbeat frequency by simulating different RR interval distributions found in the BIDMC dataset, as discussed in Section \ref{sec:simulate_RR_distribution}. To prevent overfitting, we exclude the corresponding synthetic pairs during pretraining if a pair from the real dataset is selected as a test pair.

\begin{figure}[htb!] 
\centering
\includegraphics[width=\linewidth]{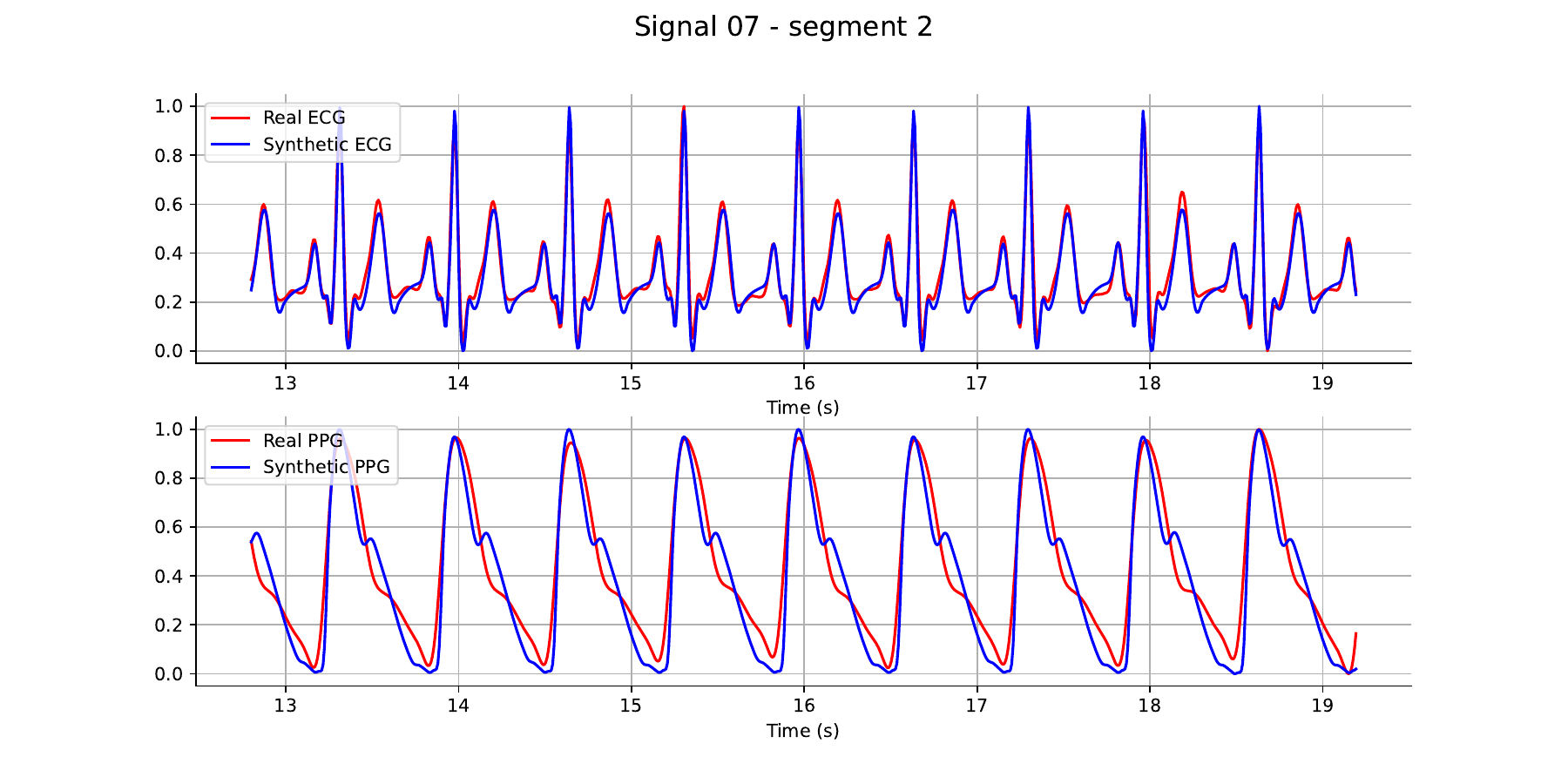}
\caption{A sample of synthetic ECG-PPG pair generated by our ODE model with parameters: 
$\bm{\theta} = \left[ -\frac{\pi}{2.3}, -\frac{\pi}{4.5}, -\frac{\pi}{6.0}, 0, \frac{\pi}{6.0}, \frac{\pi}{4.5}, \frac{\pi}{1.4}, \frac{\pi}{0.9} \right]$, $\bm{a} = \left[3.5, 2.0, -10.0, 25.0, -6.0, 2.0, 2.2, 1.0 \right]$, and $\bm{b} = \left[0.2, 0.1, 0.1, 0.15, 0.1, 0.2, 0.4, 0.4 \right]$.}
\label{fig:original07_segment2}
\end{figure}

However, some limitations of the current synthetic ECG-PPG pairs are evident in Fig. \ref{fig:original07_segment2}. Specifically, the ODE model still faces challenges in simulating small waveforms that are likely caused by noise in the ECG and PPG measurements. Additionally, in this ECG-PPG pair, positional discrepancies appear between the diastolic peaks of the clean (noise-free) synthetic PPG and the real PPG. These differences may stem from various individual influential factors, which should be further explored in future work. In the example shown in Fig. \ref{fig:original07_segment2}, we introduced three additional elements to each of the parameters $\bm{a}$, $\bm{b}$, and $\bm{\theta}$ to simulate small noisy waves between the T and P waves.

\subsubsection{ Sex-Based ECG Reconstruction Analysis}
In this section, we investigate the influence of sex on ECG reconstruction.  For our current analysis, we utilize $42$ distinct ECG-PPG pairs from the BIDMC dataset, out of which $16$ pairs belong to male subjects, while $26$ are associated with female subjects. For the male ECG reconstruction, we began by randomly selecting $2$ pairs (signal 01 and signal 24) for testing, leaving the remaining $14$ pairs for training. Training scenarios for each sex are as follows: (1) Sex-Specific Training: the model is trained exclusively on male or female data; (2) Two-Stage training: the model is first pre-trained on our synthetic dataset, then fine-tuned on sex-specific data; and (3) Mixed-Sex training: the model is trained to incorporate data from both Sexes.

\begin{figure}[htb!] 
\centering
\begin{subfigure}{0.9\textwidth}
  \centering
   \includegraphics[width=\linewidth]{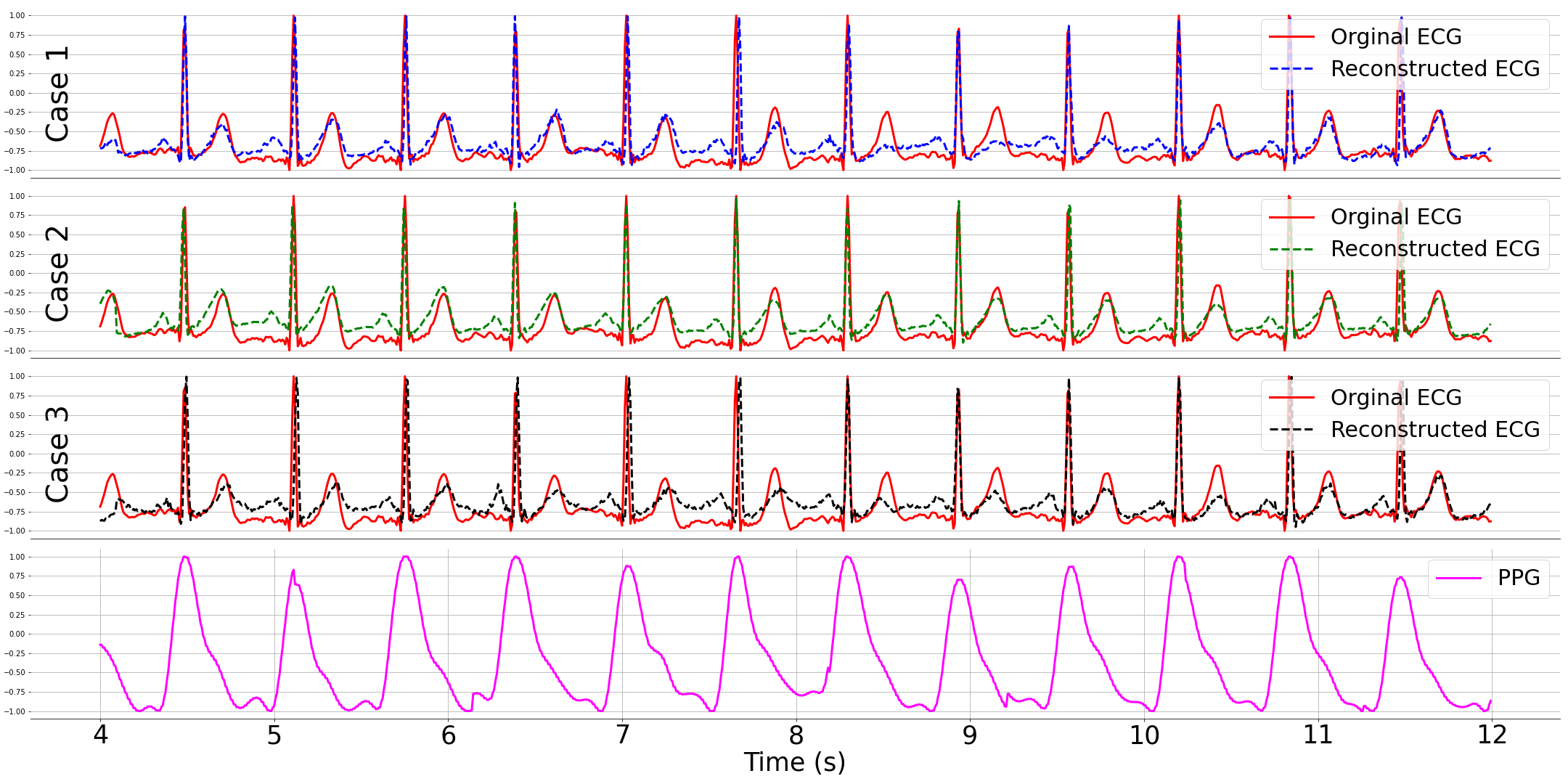}
  \caption{Signal 01.}
  \label{fig:male_01}
\end{subfigure}
\begin{subfigure}{0.9\textwidth}
  \centering
    \includegraphics[width=\linewidth]{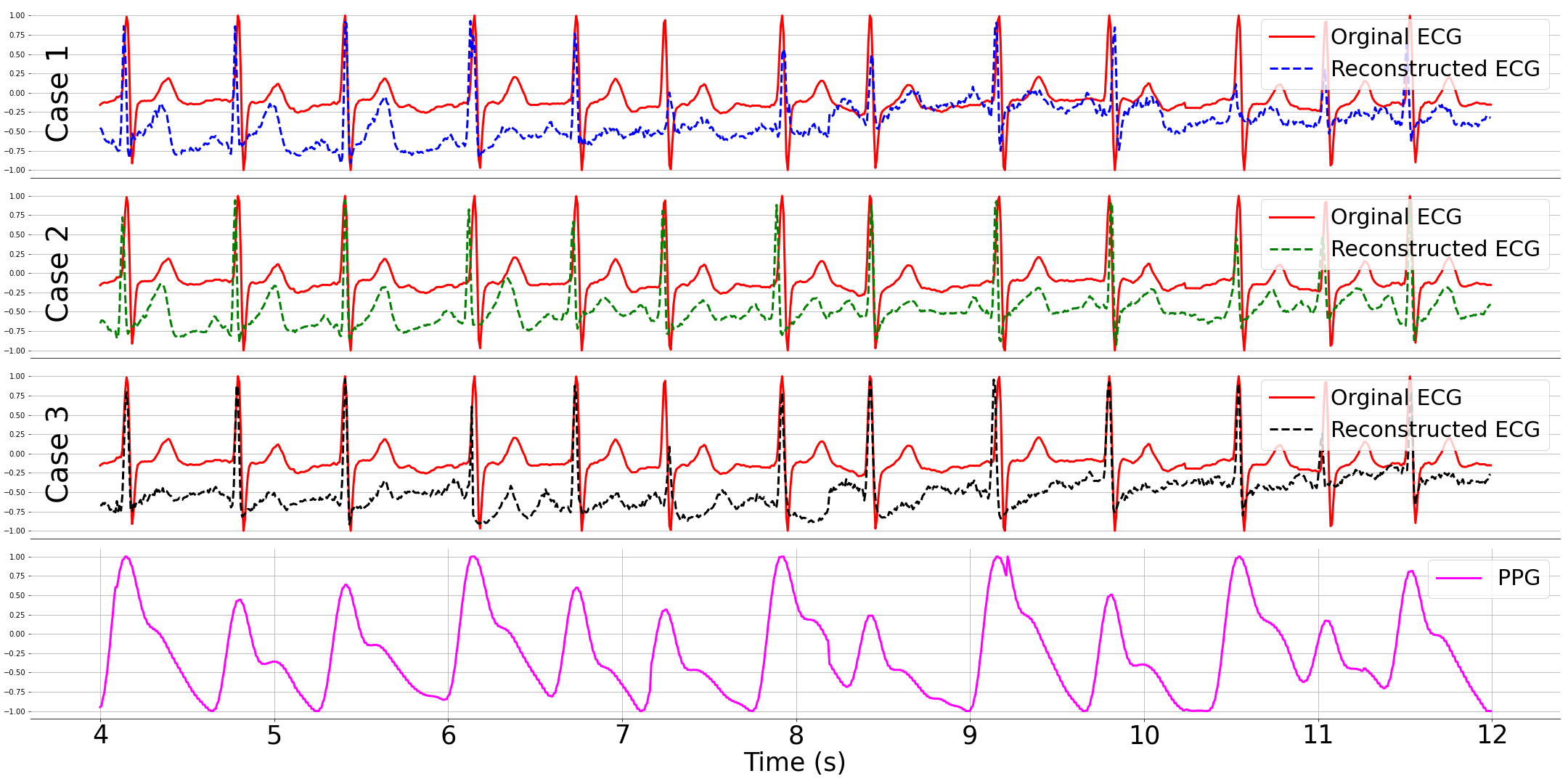}
  \caption{Signal 24.}
  \label{fig:male_24} 
\end{subfigure}
\caption{ECG reconstruction on male subjects using the proposed CLEP-GAN method: a selection of two random ECG-PPG pairs (signal 01 and signal 24) from male subjects were utilized as the testing data. Case 1 highlights the reconstructed ECGs attained when the model is exclusively trained using data from male subjects. Case 2 demonstrates the reconstructed ECG outcomes when the model undergoes a two-step training process: initial pretraining on our synthetic dataset, followed by fine-tuning utilizing data from male subjects. Case 3 illustrates the reconstructed ECG outcomes achieved when the model is trained with data encompassing both female and male subjects. }
\label{fig:male_results}.
\end{figure}

Fig. \ref{fig:male_results}  presents visualizations of reconstructed ECGs. A comparison of ECG reconstructions for signal 01 across the three training schemes reveals that training the model with mixed-sex data does not outperform the scheme that exclusively uses male subject data. The male-only training scheme achieves slightly better results, although the improvement is marginal.  Moreover, the use of transfer learning followed by fine-tuning on the male subject training set (as in case 3) produces the most accurate reconstructions, especially in the R and T waves. Reconstructing signal 24, however, remains challenging, likely due to the irregular RR interval patterns in the signal. Despite this, the transfer learning-based approach shows the most promise, delivering superior reconstruction quality. Table \ref{table:sex_results} provides quantitative results for the three schemes, clearly indicating that exclusive use of male training data yields better outcomes than mixed-sex training. Additionally, transfer learning further improves accuracy, as evidenced by the lowest RMSE and $MAE_{HR}$ values.

A similar conclusion emerges from the female experiments, as shown in Fig. \ref{fig:female_results} in the Appendix and Table \ref{table:sex_results}. Specifically, for signal 51, using both female and male training signals results in a 27\% increase in RMSE, despite a slight improvement in $MAE_{HR}$.  For the other two testing signals, employing female training data only consistently outperforms the use of a mixed-sex training dataset, demonstrating lower values for both RMSE and $MAE_{HR}$.

In conclusion, the experiments provide significant insight: sex should be considered during ECG reconstruction, as supported by findings in \cite{surawicz2003differences, ezaki2010gender, ravens2018sex, Carbone2020, Zhu2021}. Merely expanding the dataset to include both sexes may decrease performance. It is critical to recognize that reconstruction quality may be affected by subjects' health conditions, age, and signal noise. These factors could greatly differ from the characteristics of testing signals, and integrating diverse individual attributes may hinder rather than improve ECG reconstruction accuracy.

\begin{table}[htb!]
\centering
\small
\setlength{\tabcolsep}{2pt}
\begin{tabular}{@{}ccccc@{}} 
 \toprule
  Training Set & Sex & Testing Signal &  RMSE & $MAE_{HR}$ \\
 \midrule
  &   &  14 &  0.35 &  \textbf{0.74}\\ 
  &   &  30 &  0.32&  1.20\\ 
  & \multirow{-3}{*}{Female} &  51 &  \textbf{0.36} &  0.66\\ 
\cmidrule{2-5}
 &   &  01 &  0.40 &  0.97 \\ 
\multirow{-5}{*}{Female/Male} & \multirow{-2}{*}{Male}&  24 &  \textbf{0.45} &  23.98\\            
 \midrule
 &  &   14 &   \textbf{0.33} &   1.27\\ 
 &   &  30 &  \textbf{0.30} &  \textbf{0.37}\\ 
 \multirow{-3}{*}{Female/Male} & \multirow{-3}{*}{Female} &  51 &  0.37 &  0.47\\ 
\cmidrule{2-5}
 \multirow{-2}{*}{+transfer learning} &  &  01 &  \textbf{0.38} &  \textbf{0.22}\\ 
  &   \multirow{-2}{*}{Male} &  24 &  0.52 &  \textbf{3.31}\\             
 \midrule
  &    &   14 &   0.38 &   0.85\\ 
  &    &  30 &  0.41 &  0.82\\ 
  &    \multirow{-3}{*}{Female} &  51 &  0.63 &  \textbf{0.38}\\ 
\cmidrule{2-5}
  &   &  01 &  0.45 & 1.08 \\ 
  \multirow{-5}{*}{Female+Male} &   \multirow{-2}{*}{Male}  &  24 &  0.53 &  18.05 \\ 
 \bottomrule
\end{tabular}
\caption{Quantitative results of the proposed CLEP-GAN method applied to three training set conditions: (1) Training on female or male subjects alone. (2) Pretraining on our synthetic dataset followed by fine-tuning using female or male subjects. (3)Training on both female and male subjects.}
\label{table:sex_results}
\end{table}

\subsection{Ethical approval declarations}
Not applicable

\section{Discussion}
In recent research, the intricate task of unseen ECG reconstruction has gained considerable attention. Whereas the reconstruction of ECG from PPG signals isn't a novel concept, most existing studies emphasize future reference ECG, using a subset of ECG cycles for training and setting aside the rest for testing. However, unseen ECG reconstruction presents a significant challenge. This is particularly true when reconstructing smaller waves like the Q, P, S, and T waves. Moreover, several individual factors, such as sex, age, and varying health conditions, can influence the relationship between ECG and PPG signals. Simply enlarging the training set might not necessarily amplify the reconstruction quality. This leads to a substantial gap in subject-independent ECG reconstruction from PPG. Addressing this, our research introduces a novel end-to-end training framework, combining three cutting-edge techniques: contrastive learning, adversarial learning, and attention gating.

Instead of directly mapping PPG signals to corresponding ECG signals, our methodology initially involves separately training the model to reconstruct each signal type on its own, i.e., \(ECG \rightarrow ECG\) and \(PPG \rightarrow PPG\). Following this, we apply contrastive learning to the latent representations of both signals. This means we analyze and compare the deep features extracted from the ECG signals by the ECG generator network's encoder and those extracted from the PPG signals by the PPG generator network's encoder. Additionally, we integrate adversarial learning into our architecture. This involves the utilization of dual discriminators, with one focusing on the temporal characteristics and the other on the frequency components of the generated signals. 

Within our framework, we have employed two powerful models as ECG and PPG generators: the Attention U-Net and the VQ-VAE. Consequently, the frameworks based on the Attention U-Net and VQ-VAE are named ``CLEP-GAN'' and ``CLEP-VQGAN'', respectively. The strength of Attention U-Net lies in its capacity to dynamically pinpoint vital segments of the signal, thereby facilitating a more refined extraction of primary features.  Conversely, the VQ-VAE leverages the VQ technique, proficiently transforming continuous latent variables into a finite set of discrete alternatives. This transition produces a richer and smoother structure in the latent space. However, despite the merits of VQ-VAE, our empirical results indicate that the generator grounded on the Attention U-Net (CLEP-GAN) outperforms the VQ-VAE-based one (CLEP-VQGAN).

This study also introduces a pioneering approach with an ODE-based methodology for generating synthetic ECG-PPG pairs. Our ODE model employs three critical parameters: \(\bm{a}\), \(\bm{b}\), and \(\bm{\theta}\), to accurately replicate the intrinsic characteristics of the ECG signal. These parameters represent the amplitude, width, and reference angles for various wave components, specifically the P wave, Q wave, R peak, S wave, and T wave. Unlike previous ODE models that only generated ECG signals, our method is designed to produce ECG-PPG pairs, enhancing its applicability. Another innovation in our approach is the simulation of diverse ECG RR interval distributions. To effectively mimic ECG cycles, we introduce a dynamic parameter, denoted as \(f\), which reflects fluctuations in the RR intervals. This parameter \(f\) is adaptively adjusted in response to changes in the RR interval, ensuring a precise and dynamic representation of the ECG RR intervals.

To evaluate the effectiveness of the proposed methods, we began with experiments on synthetic data. The initial results indicated near-perfect ECG reconstructions. This outcome can be attributed to the inherent cleanliness of our synthetic signals and the established fixed relationship between all ECG and PPG pairs, unaffected by variables such as individual age, sex, and fluctuating health conditions.

After conducting initial tests, we proceeded to evaluate our methodologies using two real-world datasets: BIDMC and CapnoBase. In this evaluation, we compared our proposed techniques with established methods, including CardioGAN, the QRS complex-enhanced encoder-decoder, and RDDM. Among the three methodologies we proposed: the improved CardioGAN, CLEP-GAN, and CLEP-VQGAN, CLEP-GAN emerged as the most effective, outperforming the other methods across most evaluation criteria.

Moving away from a generalized training approach using extensive datasets, we focused on data sourced from individual patients. This targeted strategy is designed to minimize potential inaccuracies in ECG reconstruction that may arise from variations in age and sex. Our preliminary results indicate that training initially on our synthetic dataset and then fine-tuning the model with signals from specific subjects produces waveforms more representative of the actual ground truth. This approach outperformed direct training on a combined dataset from two larger real-world sources: BIDMC and CapnoBase. A potential explanation for this is that a larger, more diverse dataset can introduce a greater range of noise and varied ECG-PPG relationships, which may compromise the precision of ECG reconstruction.

Lastly, we focused on the impact of sex in ECG reconstruction. Our experiments, which included both female and male subjects, revealed that sex plays a crucial role in the reconstruction process. We found that indiscriminately expanding the dataset to encompass both sexes might adversely affect the accuracy of ECG reconstructions. However, it's important to acknowledge that other factors could also influence reconstruction quality. These include the subjects' health conditions, age, and the presence of noise in the signals, all of which could significantly differ from the characteristics of the signals. As we move forward, it will be vital to conduct a thorough examination of diverse datasets to further validate these preliminary findings.

\section{Conclusion}
In this study, we tackled subject-independent ECG reconstruction from PPG signals. Our innovative end-to-end training framework integrates contrastive learning, adversarial learning, and attention gating. We introduced ECG and PPG generation methods inspired by the Attention U-Net and VQ-VAE models, with empirical evaluations demonstrating the superior performance of the Attention U-Net.

A particularly novel aspect of our research was the use of an ODE-based method to generate synthetic ECG-PPG pairs. This method incorporated three key parameters for ECG characteristics, as well as a dynamic parameter for RR interval variations, ensuring an accurate representation of the ECG cycle. The ODE model successfully replicated the top three ECG rhythms.

Our initial assessments using synthetic data achieved near-perfect ECG reconstructions. When applied to real-world datasets, including BIDMC and CapnoBase, our CLEP-GAN method achieved results that were comparable to or exceeded those of existing advanced approaches. Additionally, sex was found to play a significant role in ECG reconstructions, with mixed-sex data potentially reducing accuracy. Other factors, such as age, health status, and signal quality, also warrant continued investigation.

\paragraph{Limitations and Future Work} Achieving reliable subject-independent ECG reconstruction presents numerous challenges, including noise, individual variability, and health-related differences. Developing a semantic space capable of accurately representing both observed and unobserved ECGs remains a significant challenge, largely due to the need for diverse real-world data. Synthetic data generation offers a resource-efficient alternative, though replicating the full range of real-world scenarios with synthetic data introduces additional complexities. Our ODE-based approach for generating synthetic ECG-PPG pairs provides a promising avenue, though challenges remain, particularly in simulating smaller waveforms that may be affected by noise in ECG measurements. Modifying the ODE-based model to better capture these smaller waveforms would enable the neural network model to predict finer waveform details, and this should be a priority in future work. Additionally, while our CLEP-GAN method demonstrates improved overall performance compared to other approaches, precise reconstruction of smaller waves, such as the T wave, remains an area for enhancement. In future work, addressing these nuances will be a primary focus as we continue to explore the complexities of ECG waveform reconstruction from PPG signals.

\backmatter


\section*{Declarations}
\subsection{Funding}
This study was funded by the Natural Sciences and Engineering Research Council of Canada (NSERC).
\subsection{Conflict of interest}
On behalf of all the authors of this work, we hereby declare that no conflicts of interest have been identified for this research.
\subsection{Ethics approval}
Not applicable

\subsection{Consent to participate}
Not applicable

\subsection{Consent to publication}
Not applicable

\subsection{Availability of data and materials}
Not applicable

\subsection{Code availability}
Please refer to this link: https://github.com/Mathematics-Analytics-Data-Science-Lab/CLEP-GAN.git

\subsection{Authors’ contributions}
\begin{itemize}
    \item Conceptualization and methodology: X.L. and H.H.
    \item Code: X.L. and N.A.
    \item Writing original draft: X.L.
    \item Writing review and editing: X.L., H.H., S.X., and F.H.
    \item Supervision: A.G. and H.H.
    \item Final review: All authors
\end{itemize}

\clearpage
\bibliography{sn-article}

\clearpage
\begin{appendices}
\section{Adversarial Learning} \label{sec:gan}
Within the realm of generative modelling, GANs have emerged as a pioneering framework, enabling the generation of high-quality data that closely resembles real-world samples. GANs consist of two key components: a generator and a discriminator. The generator crafts synthetic data samples, while the discriminator evaluates whether a given sample is real or generated. Through an adversarial process, the generator refines its outputs to become progressively more convincing, while the discriminator enhances its ability to differentiate between real and synthetic data. This dynamic interplay interplay drives continuous improvement, ultimately resulting in the generation of data that aligns closely with real observations. In our pursuit to enhance the ECG reconstruction process, we harness the potency of GAN-based adversarial learning. By integrating GAN architecture into our generator network, we empower it to yield reconstructed ECG signals with high accuracy.

To maintain fidelity to both time and frequency characteristics of cardiac dynamics, we adopt a dual discriminator strategy \cite{Nguyen2017, Sarkar2021}. To utilize a frequency-domain-based discriminator, we first need to transform the signals from the time domain to the frequency domain. This approach involves employing the Short-Time Fourier Transformation (STFT) on the ECG and PPG time series data. Denoting the time series as $x[n]$, the STFT on $x[n]$, denoted as $STFT\{x[n]\}(m, \omega) \equiv X(m, \omega)=\sum_{n=-\infty}^{\infty}x[n]w[n-m]e^{-i\omega n}$, captures the data's spectral content, where $m$ signifies the step size and $w[n]$ represents the Hann window function. The spectrogram is ultimately derived from $STFT_{spect}(x[n]) = \log(|X(m, \omega)|+\delta)$, where $\delta$ is a small number added to avoid potential infinite conditions.

\begin{figure}[htb!] 
\centering
\begin{subfigure}{0.9\textwidth}
  \centering
  \includegraphics[width=\linewidth]{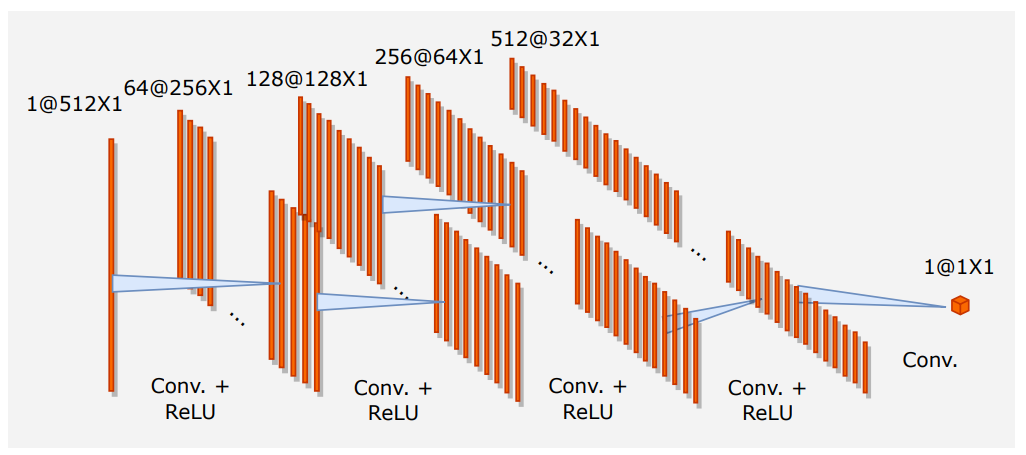}
  \caption{Time domain-based discriminator.}
  \label{fig:Discriminator_time}
\end{subfigure}
\begin{subfigure}{0.9\textwidth}
  \centering
  \includegraphics[width=\linewidth]{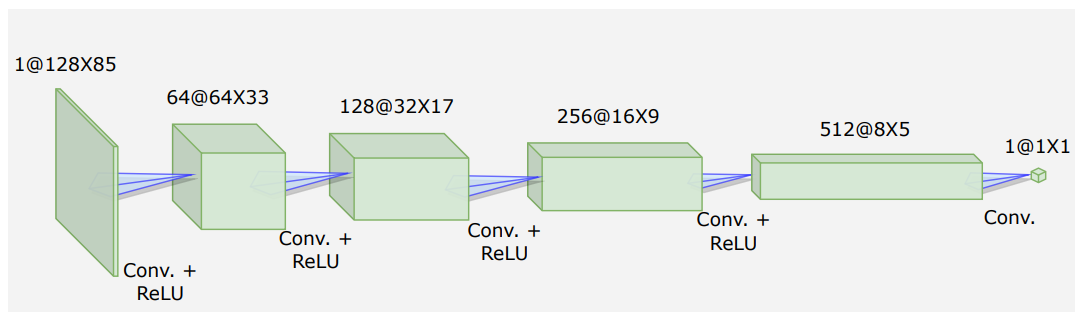}
  \caption{Frequency domain-based discriminator.}
  \label{fig:Discriminator_freq}
\end{subfigure}
\caption{Architectures of dual discriminators.  The format ``C@HxW'' describes the dimensions of a feature layer. ``C'' stands for the number of channels, ``H'' indicates the height of the feature map, and ``W'' specifies its width. For instance, ``1@512x1'' refers to a layer with a single channel, a height of $512$, and a width of $1$.}
\label{fig:Discriminators}.
\end{figure}

\section{Attention U-Net} \label{sec:Attention_UNet}
\begin{figure}[htb!] 
\centering
\includegraphics[width=0.9\linewidth]{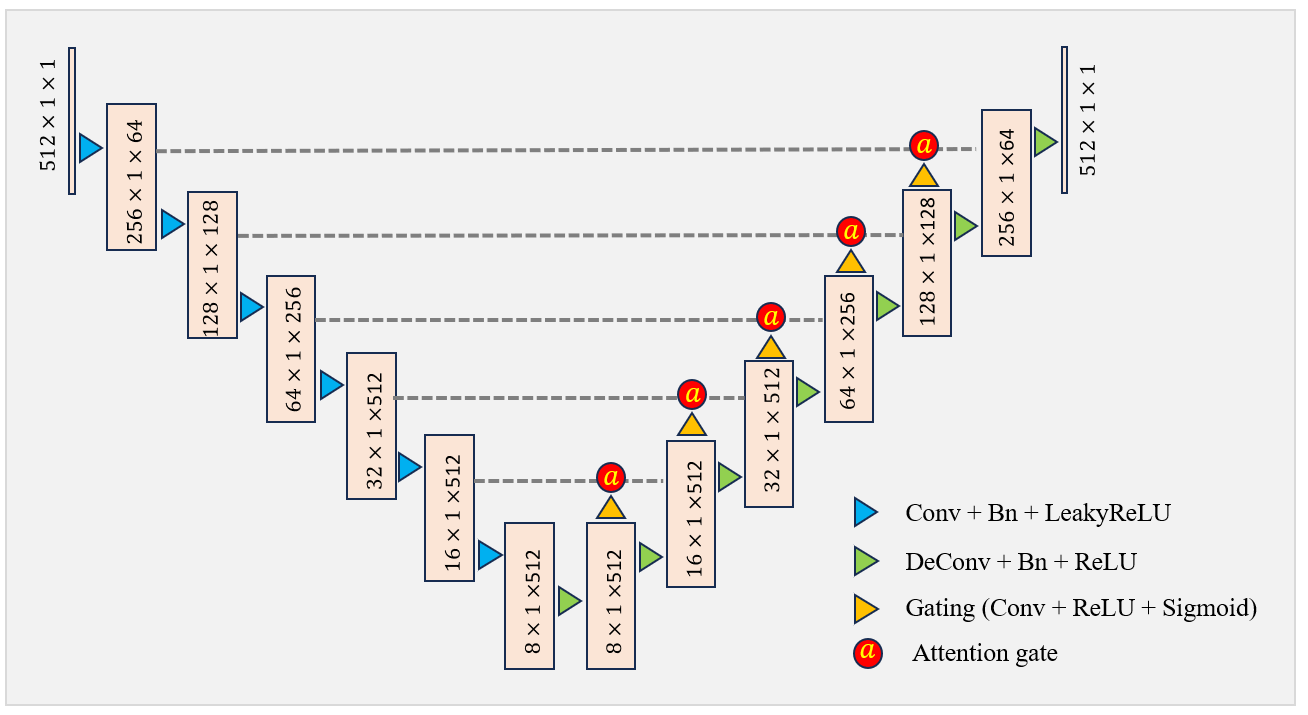}
\caption{The architecture of Attention U-Net generator.}
\label{fig:Attention_UNet}
\end{figure}

The Attention U-Net incorporates an attention gate (AG) into the standard U-Net architecture. Fig. \ref{fig:Attention_UNet} illustrates the architecture of the Attention U-Net, emphasizing its capacity to concentrate on vital features within the skip connections. Within this structure, features denoted by \(x_i^l\) are sourced from the skip connection at layer \(l\). The gating vector, symbolized by \(g\), designates the focal region. These features and the gating vector are mapped to an intermediate-dimensional space \(R^{F_{\text{int}}}\), with \(F_{\text{int}}\) specifying the dimensions of this space. The objective is to derive scalar attention values, \(\alpha^l_i\), for each temporal unit \(x_i^l \in R^{F_l}\) based on the gating vector \(g_i \in R^{F_g}\).

To achieve this, linear transformations are applied to $x_i^l$ and $g_i$ using weights and biases, denoted as $\theta_x = W_x x_i^l + b_x$ and $\theta_g = W_g g_i + b_g$, respectively. Here, $W_x \in R^{F_l \times F_{\text{int}}}$, $W_g \in R^{F_g \times F_{\text{int}}}$, and $b_x$, $b_g$ are the bias terms. Following these transformations, a non-linear ReLU activation (denoted as $\sigma_1$) is applied to yield a summed feature activation $f = \sigma_1(\theta_x + \theta_g)$.

Subsequently, a linear mapping of $f$ onto the $R^{F_{\text{int}}}$ dimensional space occurs through channel-wise $1 \times 1$ convolutions. The result is passed through a sigmoid activation function ($\sigma_2$) to obtain attention weights. The attention map for $x_i^l$ is derived as $\alpha^l_i = \sigma_2(\psi * f)$, where $\psi \in R^{F_{\text{int}}}$ and * denotes convolution. Finally, we perform element-wise multiplication between $x_i^l$ and $\alpha^l_i$ to yield the ultimate output from the attention layer.

\section{Discrete Latent variables} \label{sec:discrete_latent_variables}
The authors in \cite{VQVAE2017} introduced the concept of VQ-VAE, which employs discrete latent variables and utilizes a novel training methodology inspired by vector quantization (VQ). In this structure, both posterior and prior distributions are categorical, and the samples extracted from these distributions are used to index an embedding table. These indexed embeddings subsequently serve as inputs for the decoder network.

A latent embedding space is delineated as \(e \in \mathbb{R}^{K \times D}\), where \(K\) signifies the size of the discrete latent space, represented as a \(K\)-way categorical variable, and \(D\) denotes the dimensionality of each latent embedding vector \(e_i\). It should be noted that there are \(K\) embedding vectors denoted by \(e_i \in \mathbb{R}^{D}\), where \(i = 1, 2, \ldots, K\). The model accepts an input \(x\) that traverses through an encoder, yielding an output \(z_e(x)\). Following this, the discrete latent variables \(z\) are computed using a nearest-neighbor lookup mechanism that employs the shared embedding space \(e\), as illustrated in the posterior categorical distribution \(q(z|x)\) probabilities:

\begin{align}
q(z=k|x) 
= \begin{cases} 
    1 & \text{if } k = \text{argmin}_i\|z_e(x) - e_i\|_2,\\
    0 & \text{otherwise.}
   \end{cases}
\end{align}

In this scenario, a singular random variable \(z\) is utilized to represent the discrete latent variables for simplicity. The decoder receives the respective embedding vector \(e_k\) as denoted by \(z_q(x) = e_k\), where \(k = \text{argmin}_j \|z_e(x) - e_j\|_2\). This forward computation pathway can be perceived as a standard autoencoder, characterized by a unique non-linearity that aligns the latent variables to one of the \(K\) embedding vectors. Viewing this model through the lens of a VAE, it is observed that \(q(z=k|x)\) operates deterministically. Furthermore, by establishing a uniform prior over \(z\), the KL divergence remains consistent, equating to \(\log K\).

\begin{figure}[htb!] 
\centering
\includegraphics[width=0.9\linewidth]{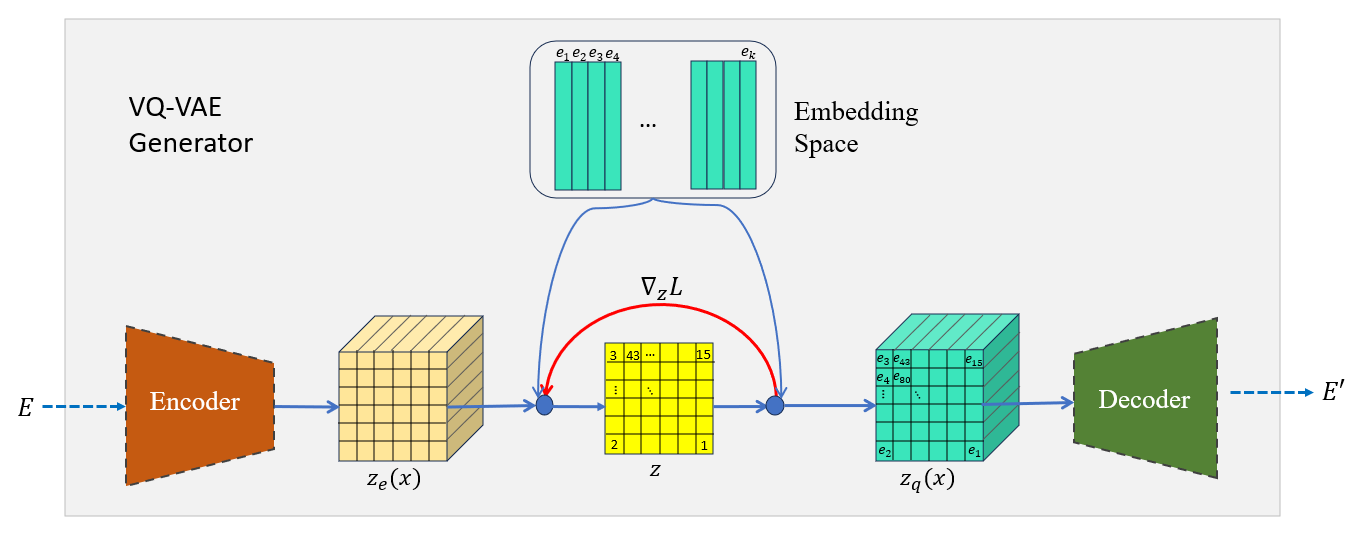}
\caption{The architecture of VQ-VAE. The encoder output, denoted as $z_{e}(x)$, is mapped to the nearest points, $e_i$, within the embedding space. During forward computation, the nearest embedding, $z_{q}(x)$, is passed to the decoder, and during the backward pass the gradient, $\nabla_{z}L$, is transmitted unaltered back to the encoder.}
\label{fig:VQVAE}
\end{figure}

\begin{figure}[htb!] 
\centering
\begin{subfigure}{0.9\textwidth}
  \centering
  \includegraphics[width=\linewidth]{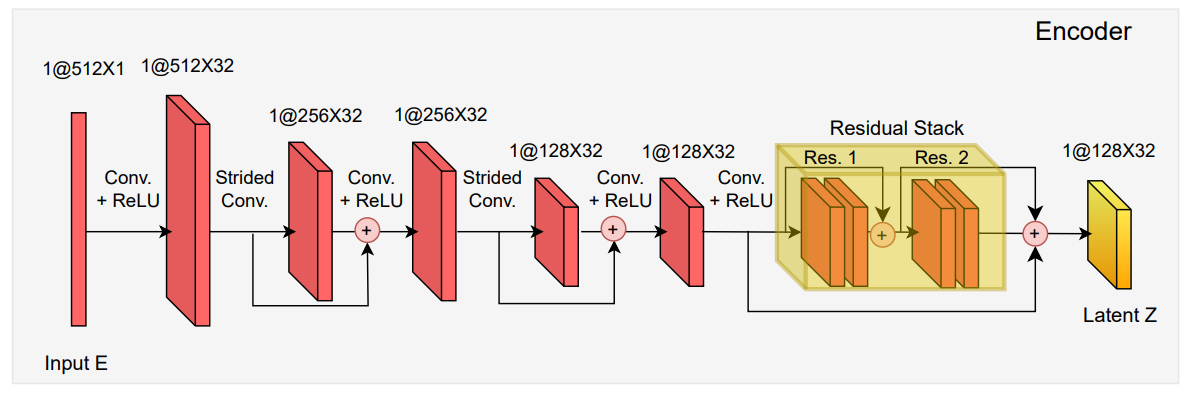}
  \caption{Encoder of VQ-VAE.}
  \label{fig:VQVAE_encoder}
\end{subfigure}
\begin{subfigure}{0.9\textwidth}
  \centering
  \includegraphics[width=\linewidth]{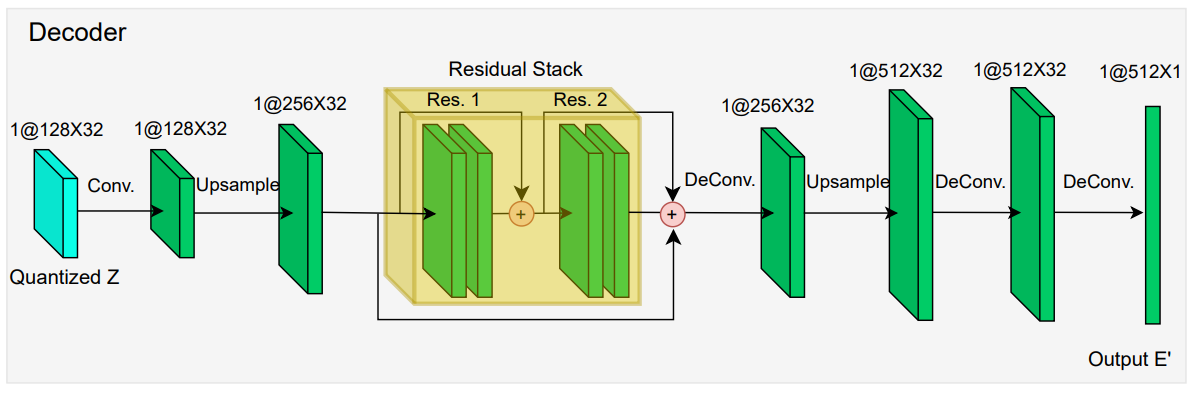}
  \caption{Decoder of VQ-VAE.}
  \label{fig:VQVAE_decoder}
\end{subfigure}
\caption{A depiction of the VQ-VAE's encoder and decoder architectures. Both the encoder and decoder utilize residual stacks, each comprising two residual blocks.}
\label{fig:VQVAE_encoder_decoder}.
\end{figure}

\section{Inspecting the Deviation of RR Intervals} As previously discussed, when using RMSE as the evaluation metric, the performance of the original CardioGAN lags behind that of other methods. Additionally, Fig. \ref{fig:reconstructed_real_ECGs} illustrates that it faces more challenges in reconstructing small waveforms compared to other methods. However, as indicated in Table \ref{table:results_HRV}, CardioGAN showcases impressive accuracy in HRV, surpassing all other methods.

\begin{figure}[htb!] 
\centering
\includegraphics[width=0.6\linewidth]{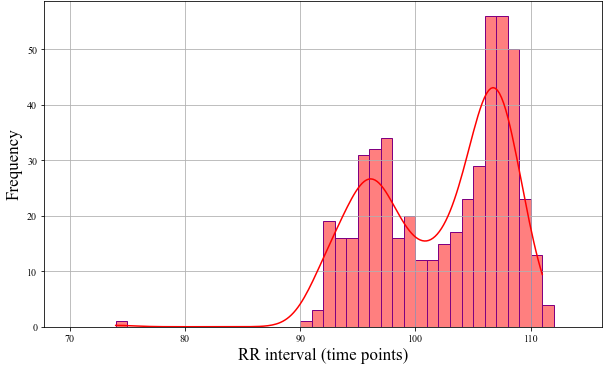}
\caption{RR distribution of signal 0332 in the CapnoBase dataset. RR intervals are measured in data points.}
\label{fig:RR_0332}
\end{figure}

\begin{figure}[htb!] 
\centering
\includegraphics[width=\linewidth]{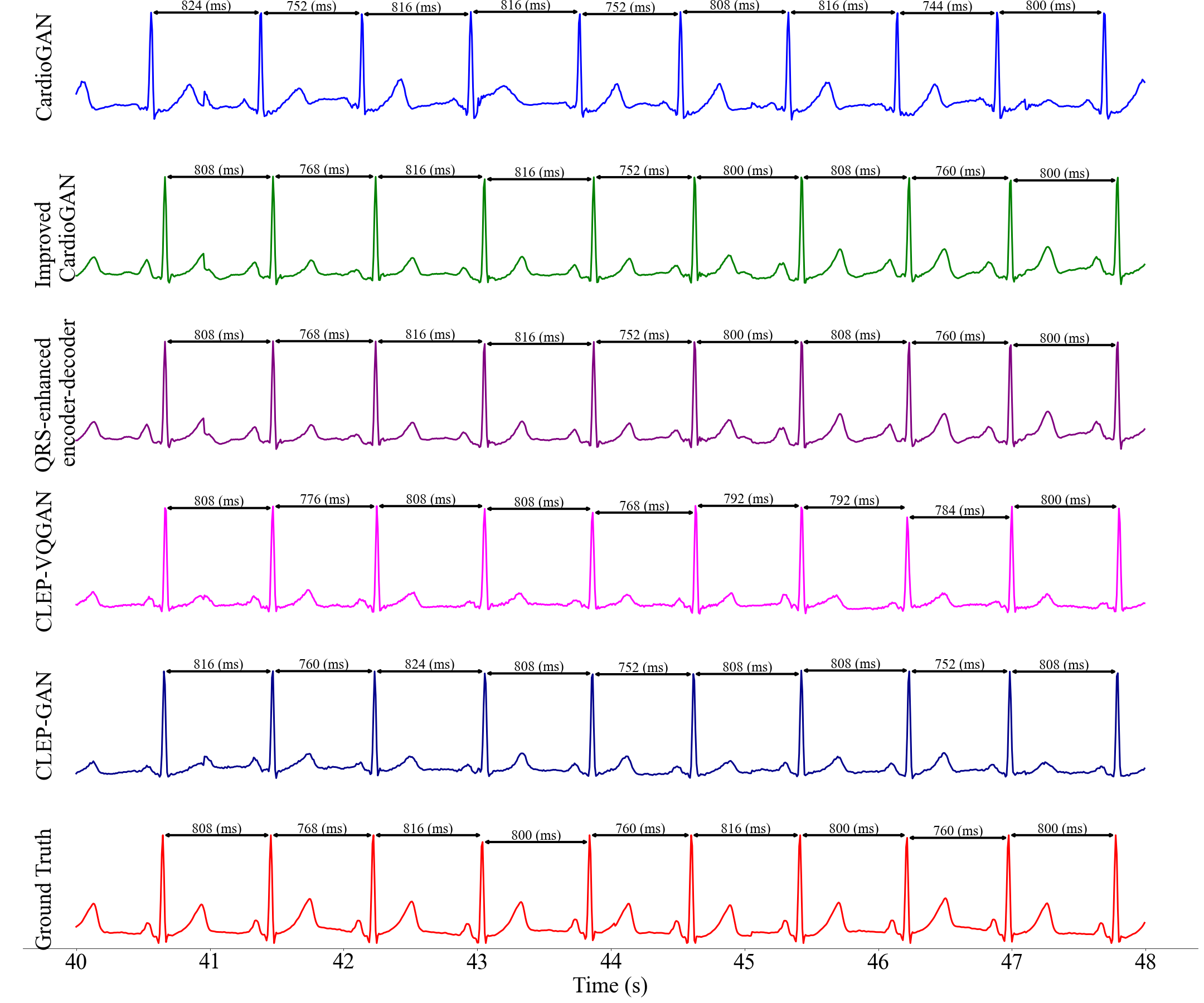}
\caption{Cycles of the reconstructed signal 0332 from the CapnoBase dataset labeled with RR intervals. ECGs were reconstructed using five methods: three proposed (improved CardioGAN, CLEP-VQGAN, and CLEP-GAN) and two advanced (CardioGAN \cite{Sarkar2021} and QRS complex-enhanced encoder-decoder \cite{Chiu2020}).}
\label{fig:RR_five_methods}
\end{figure}

To visualize the RR intervals, we have plotted the ground truth RR interval distribution of signal 0332 in CapnoBase, as shown in Fig. \ref{fig:RR_0332}. We have also included several cycles of reconstructed ECGs labeled with corresponding RR intervals from five different methods, as displayed in Fig. \ref{fig:RR_five_methods}.

From Fig. \ref{fig:RR_0332}, it is evident that the RR interval distribution of signal 0332 exhibits a bimodal shape, resembling two connected Gaussian distributions. However, there are several RR intervals significantly distant from the left Gaussian-like distribution, resulting in a large standard deviation of the RR intervals.

Analyzing Fig. \ref{fig:RR_five_methods}, we observe that the reconstructed ECGs using CardioGAN consistently exhibit a leftward shift, leading to a higher RMSE. An interesting observation is that, although each RR interval does not exactly match the corresponding ground truth values, the model attempts to balance the RR interval distribution. For instance, in the case of CardioGAN, when the first RR interval of the reconstructed ECG is $824$ ms, whereas its ground truth has a smaller interval of $808$ ms, the second RR interval of the reconstructed ECG becomes $752$ ms, which is $16$ ms shorter than its ground truth, in an attempt to achieve balance. This balancing effect is noticeable in the reconstructed ECGs of other methods as well.

\begin{figure}[htb!] 
\centering
\includegraphics[width=0.6\linewidth]{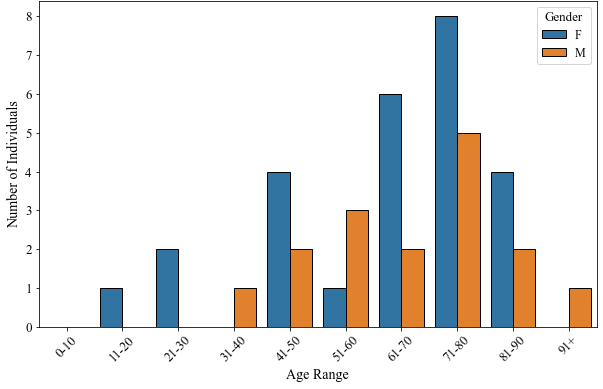}
\caption{Illustration of the sex and age distribution of subjects in the BIDMC Dataset. We utilized $42$ distinct ECG-PPG pairs from the BIDMC dataset.}
\label{fig:sex_age_dist}
\end{figure}

The sex and age distribution of the ECG-PPG pairs used in our experiments is depicted in Fig. \ref{fig:sex_age_dist}.

\begin{figure}[htb!] 
\centering
\begin{subfigure}{0.7\textwidth}
  \centering
   \includegraphics[width=\linewidth]{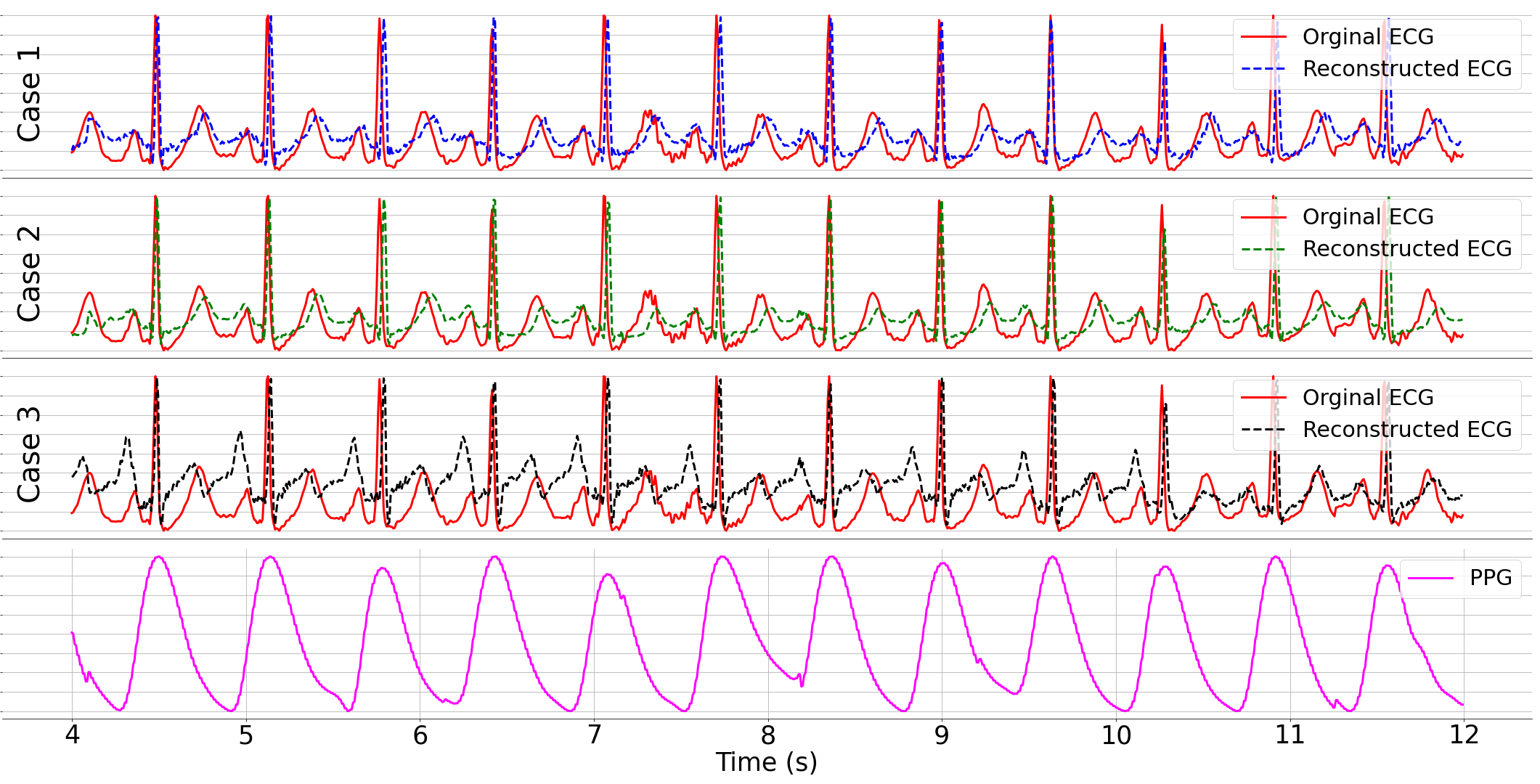}
  \caption{Signal 14.}
  \label{fig:female_14}
\end{subfigure}
\begin{subfigure}{0.7\textwidth}
  \centering
    \includegraphics[width=\linewidth]{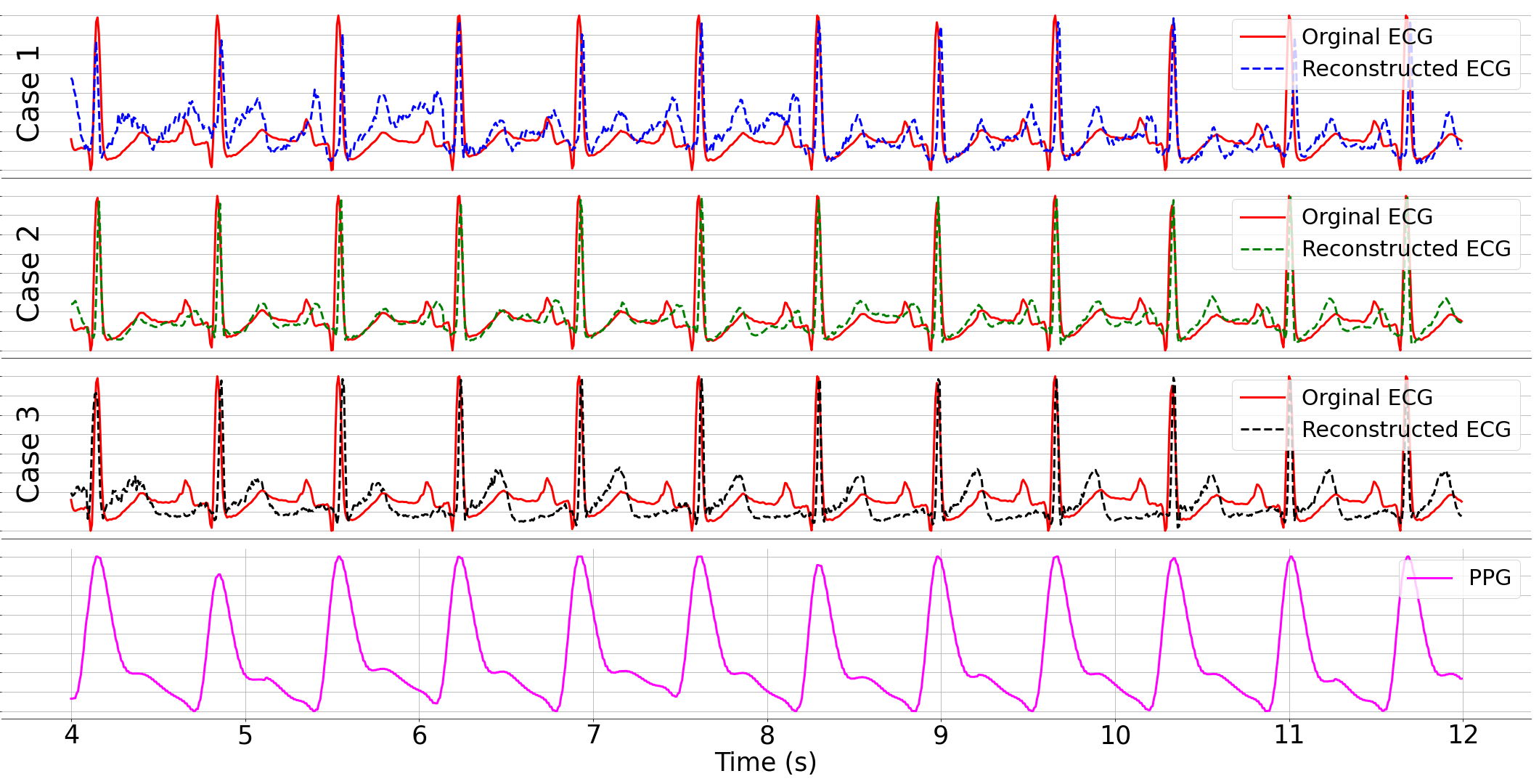}
  \caption{Signal 30.}
  \label{fig:female_30} 
\end{subfigure}
\begin{subfigure}{0.7\textwidth}
  \centering
    \includegraphics[width=\linewidth]{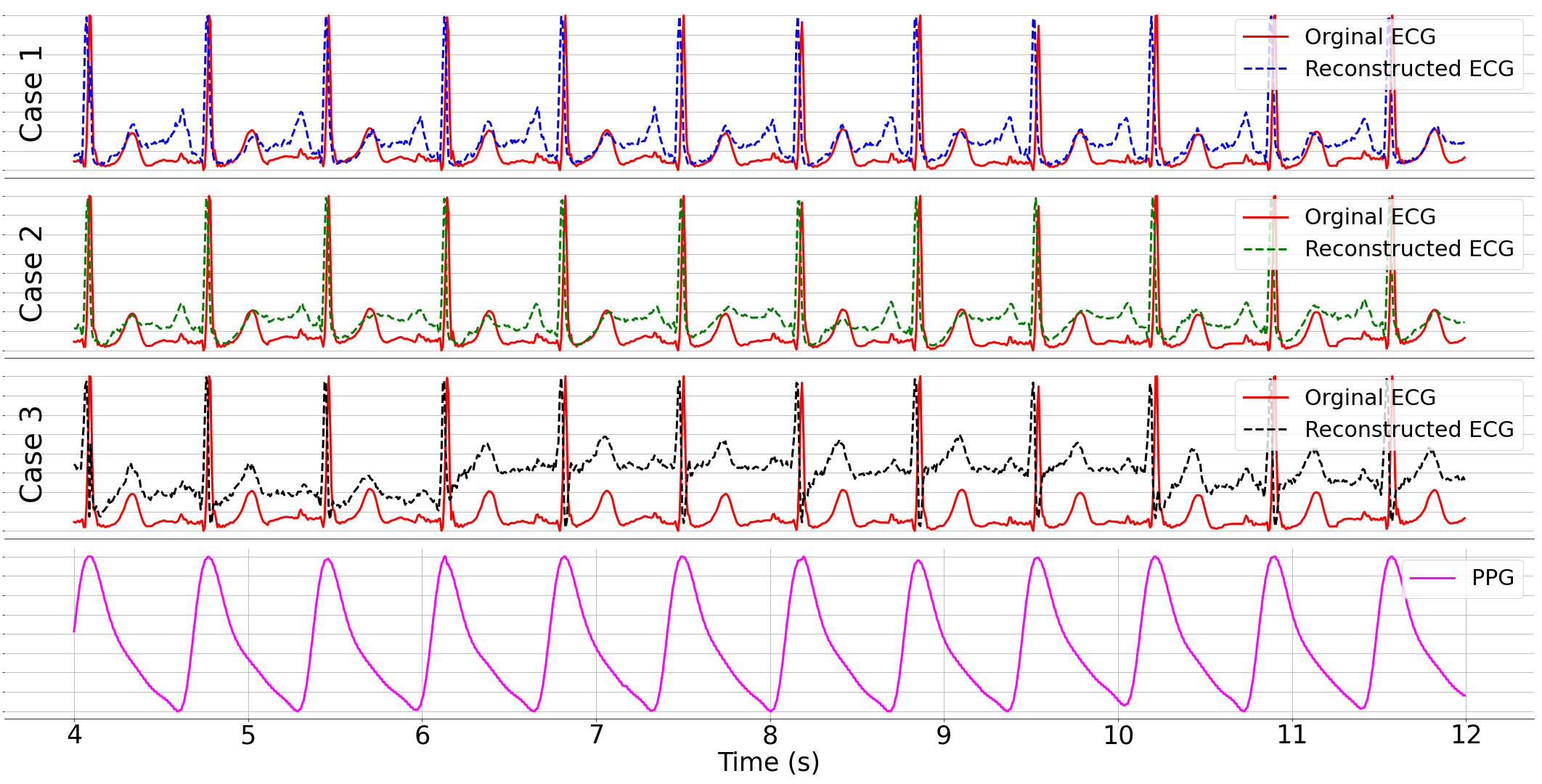}
  \caption{Signal 51.}
  \label{fig:female_51}
  \end{subfigure}
\caption{ECG reconstruction on female subjects using the proposed CLEP-GAN method: a selection of three random ECG-PPG pairs from female subjects were used as the testing data. Case 1 highlights reconstructed ECGs achieved when exclusively training the model using data from female subjects. Case 2 demonstrates the reconstructed ECGs when the model undergoes a two-step process: initial pretraining on our synthetic dataset followed by fine-tuning exclusively with data from female subjects. Case 3 illustrates the reconstructed ECGs obtained when the model is trained utilizing data from both female and male subjects.}
\label{fig:female_results}.
\end{figure}




\end{appendices}

\end{document}